\documentclass[aps,prc,showpacs,10pt,floatfix,nofootinbib,twocolumn]{revtex4}

\newcommand{\ifproofpre}[2]{#1}

\usepackage{graphicx}
\usepackage{amsmath}
\usepackage{amssymb}
\usepackage{mathtools}
\usepackage{isotope}
\usepackage{dcolumn}
\usepackage{dsfont} 

\usepackage{wignert}
\usepackage{mcmath}



\newcommand{\Nmax}{{N_\text{max}}}
\newcommand{\Ntot}{{N_\text{tot}}}

\newcommand{\Trel}{{T_\text{rel}}}

\newcommand{\bho}{b_\text{HO}}   
\newcommand{\bcs}{b_\text{CS}}   
\newcommand{\Ncm}[1][]{{N_\text{c.m.}^{#1}}}  
  
\newcommand{\rrel}{r_\text{rel}} 
\newcommand{\rrelpp}{r_{\text{rel},pp}} 
\newcommand{\rrelpn}{r_{\text{rel},pn}} 
\newcommand{\rrelnn}{r_{\text{rel},nn}} 

\newcommand{\Ncut}{{N_\text{cut}}}  

\newcommand{\wfho}{\Psi}

\newcommand{\MeV}{{\mathrm{MeV}}}
\newcommand{\fm}{{\mathrm{fm}}}

\newcommand{\hw}{{\hbar\Omega}}

\newcommand{\hwint}{{\hbar\Omega_{\text{int}}}}

\begin{document}


\title{\boldmath Halo nuclei $\isotope[6]{He}$ and $\isotope[8]{He}$ with the Coulomb-Sturmian basis}

\author{M. A. Caprio}
\affiliation{Department of Physics, University of Notre Dame, Notre Dame, Indiana 46556-5670, USA}

\author{P. Maris}
\affiliation{Department of Physics and Astronomy, Iowa State University, Ames, Iowa 50011-3160, USA}

\author{J. P. Vary}
\affiliation{Department of Physics and Astronomy, Iowa State University, Ames, Iowa 50011-3160, USA}

\date{\today}

\begin{abstract}
The rapid Gaussian falloff of the oscillator functions at large radius
makes them poorly suited for the description of the asymptotic
properties of the nuclear wave function, a problem which becomes
particularly acute for halo nuclei. We consider an alternative basis
for \textit{ab initio} no-core configuration interaction (NCCI)
calculations, built from Coulomb-Sturmian radial functions, allowing
for realistic exponential falloff.  NCCI calculations are carried out
for the neutron halo nuclei $\isotope[6,8]{He}$, as well as the
baseline case $\isotope[4]{He}$, with the JISP16 nucleon-nucleon
interaction.  Estimates are made for the root-mean-square radii of the
proton and matter distributions.
\end{abstract}


\maketitle

\section{Introduction}
\label{sec-intro}

The \textit{ab initio} theoretical
description of light nuclei is based on direct solution of the nuclear
many-body problem given realistic nucleon-nucleon
interactions.
In no-core configuration interaction (NCCI)
calculations~\cite{navratil2000:12c-ab-initio,navratil2000:12c-ncsm,barrett2013:ncsm},
the nuclear many-body problem is formulated as a
matrix eigenproblem.  The Hamiltonian is represented in terms of basis states which
are antisymmetrized products of single-particle states for the full
$A$-body system of nucleons, \textit{i.e.}, with no assumption of an
inert core.  

In practice, the nuclear many-body calculation must be carried out in
a truncated space.  The dimension of the problem grows combinatorially
with the size of the included single-particle space and with the
number of nucleons in the system.  Computational restrictions
therefore limit the extent to which converged results can be obtained,
for energies or for other properties of the wave functions.  Except
for the very lightest systems ($A\lesssim 4$), convergence is
generally beyond reach.  Based on the still-unconverged calculations
which are computationally feasible, we seek to obtain a reliable
estimate of the true values of observables which would be obtained in
the full, untruncated space.  Improved accuracy may be pursued both
through the development of bases which yield accelerated convergence,
as considered here, and by developing means by which robust
extrapolations can be
made~\cite{forssen2008:ncsm-sequences,bogner2008:ncsm-converg-2N,maris2009:ncfc,coon2012:nscm-ho-regulator,furnstahl2012:ho-extrapolation,more2013:ir-extrapolation,jurgenson2013:ncsm-srg-pshell}.

A prominent feature in light nuclei is the emergence of
halo structure~\cite{jonson2004:light-dripline,tanihata2013:halo-expt}, in
which one or more loosely-bound nucleons surround a compact core,
spending much of their time in the classically-forbidden region.  A
realistic treatment of the long-range properties of the
wave function has been found to be essential for an accurate reproduction of the halo
structure~\cite{quaglioni2009:ncsm-rgm}.

However, NCCI calculations have so far been based almost
exclusively upon bases constructed from harmonic oscillator
single-particle wave functions.  The harmonic oscillator radial
functions have the significant limitation that they display Gaussian asymptotic
behavior, \textit{i.e.}, falling as $e^{-\alpha r^2}$ for
large $r$.  The actual asymptotics for nucleons bound by a
finite-range force are instead expected to be exponential,
\textit{i.e.}, falling as $e^{-\beta r}$.  

Observables which are sensitive to the large-distance asymptotic
portions of the nuclear wave function present a special challenge to
convergence in NCCI calculations with a conventional oscillator basis.
Such ``long-range'' observables include the root-mean-square (RMS)
radius~--- an essential observable for halo nuclei~--- and $E2$
moments and transitions.  The $r^2$ dependence of the relevant
operators in both cases preferentially weights the larger-$r$ portions
of the wave function.  The results for these observables in NCCI
calculations are in general highly
basis-dependent~\cite{bogner2008:ncsm-converg-2N,cockrell2012:li-ncfc,maris2013:ncsm-pshell}.
\begin{figure}
\centerline{\includegraphics[width=\ifproofpre{1}{0.6}\hsize]{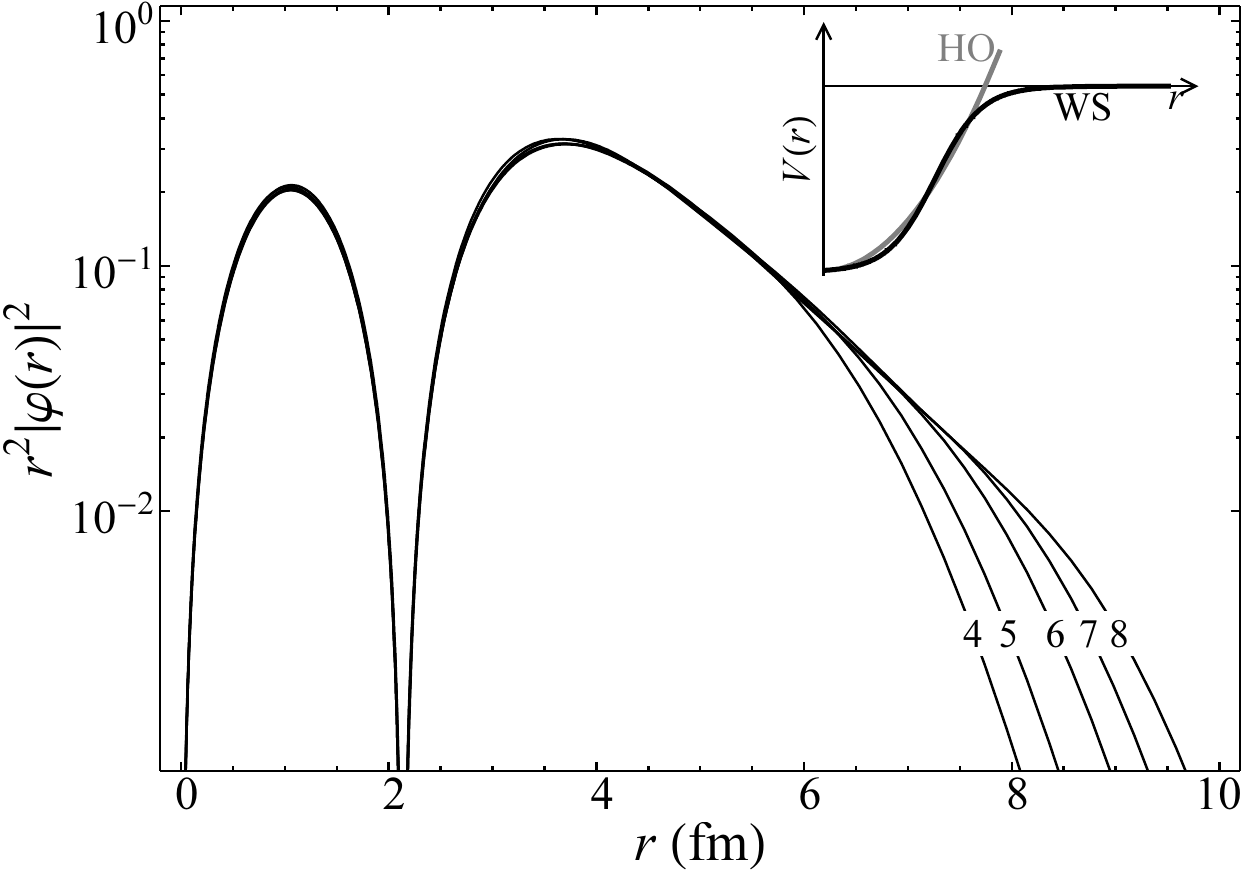}}
\caption{The calculated wavefunction obtained when a
problem with exponential asymptotics~--- here, the Woods-Saxon problem
is taken
for illustration~--- is solved in a finite basis of
oscillator functions.  The radial probability density $r^2\abs{\varphi(r)}^2$ is shown on a
logarithmic scale, so that exponential asymptotics would appear as a
straight line.  The
Woods-Saxon (WS) and harmonic oscillator (HO) potentials are shown in the inset.
(Solutions are for the Woods-Saxon $1s_{1/2}$  function, with  potential parameters appropriate to neutrons in
$\isotope[16]{O}$~\cite{suhonen2007:nucleons-nucleus}, with maximal
basis radial quantum numbers $n$ as
indicated.)  
}
\label{fig-ws-soln}      
\end{figure}

The difficulties encountered in using an oscillator basis to describe a
system with exponential asymptotics may be illustrated through the
simple one-dimensional example of the Schr\"odinger equation with a
Woods-Saxon potential.  In Fig.~\ref{fig-ws-soln}, we see the results
of solving for a particular eigenfunction in terms of successively
larger bases of oscillator radial functions.  In the classically
forbidden region, where the potential is nearly flat, the tail of the
wave function should be exponential.  It should thus appear as a
straight line on the logarithmic scale in Fig.~\ref{fig-ws-soln}.
Inclusion of each additional basis function yields a small extension
to the region in which the expected straight-line behavior is
reproduced.  However, for any finite number of oscillator functions, there
is a radius beyond which the calculated tail is seen to sharply fall
below the true asymptotics.

We are therefore motivated to consider alternative bases which might
be better suited for expanding the nuclear wave function in its
asymptotic region.  The Coulomb-Sturmian
functions~\cite{weniger1985:fourier-plane-wave}, which are obtained as
solutions of the Sturm-Liouville problem associated with the Coulomb
potential, constitute a complete set of square-integrable functions
with exponential asymptotics.  These functions have previously been
applied to few-body problems in
atomic~\cite{shull1955-continuum,rotenberg1962:sturmian-scatt,rotenberg1970:sturmian-scatt},
hadronic~\cite{jacobs1986:heavy-quark-sturmian,keister1997:on-basis,pervin2005:diss},
and nuclear~\cite{smith2012:fishbone-sturmian} physics.  The framework
for carrying out NCCI calculations with general radial basis
functions~--- and with the Coulomb-Sturmian functions, in particular~--- has been
developed in Ref.~\cite{caprio2012:csbasis}.

In the present work, we apply the Colomb-Sturmian basis to NCCI
calculations for the lightest neutron halo nuclei~--- $\isotope[6,8]{He}$~--- as
well as to the baseline case $\isotope[4]{He}$, for which converged
results can be obtained.  
Motivated by the disparity between proton and neutron radial
distributions in the neutron-rich halo nuclei, we explore the use of
proton-neutron asymmetric bases, with different length scales for the proton and neutron radial basis
functions.  We also examine the possibility of extracting
RMS radii for the proton and matter distributions based on a
relatively straightforward estimate, the ``crossover
point''~\cite{bogner2008:ncsm-converg-2N,cockrell2012:li-ncfc}, 
pending further development of more sophisticated
extrapolation schemes~\cite{coon2013:ncsm-extrapolation-ntse13-DUMMY,furnstahl2014:ir-expansion}.  The bases and methods are first reviewed
(Sec.~\ref{sec-methods}), after which the results for
$\isotope[4,6,8]{He}$ are discussed and compared with experiment (Sec.~\ref{sec-results}).
Details of the calculation of RMS radii for general single-particle bases are given in the Appendix.
Preliminary results were reported in Ref.~\cite{caprio2013:cshalo-ntse13}.

\section{Basis and methods}
\label{sec-methods}

\subsection{Basis functions}
\label{sec-methods-basis}
\begin{figure}
\centerline{\includegraphics[width=\ifproofpre{0.8}{0.4}\hsize]{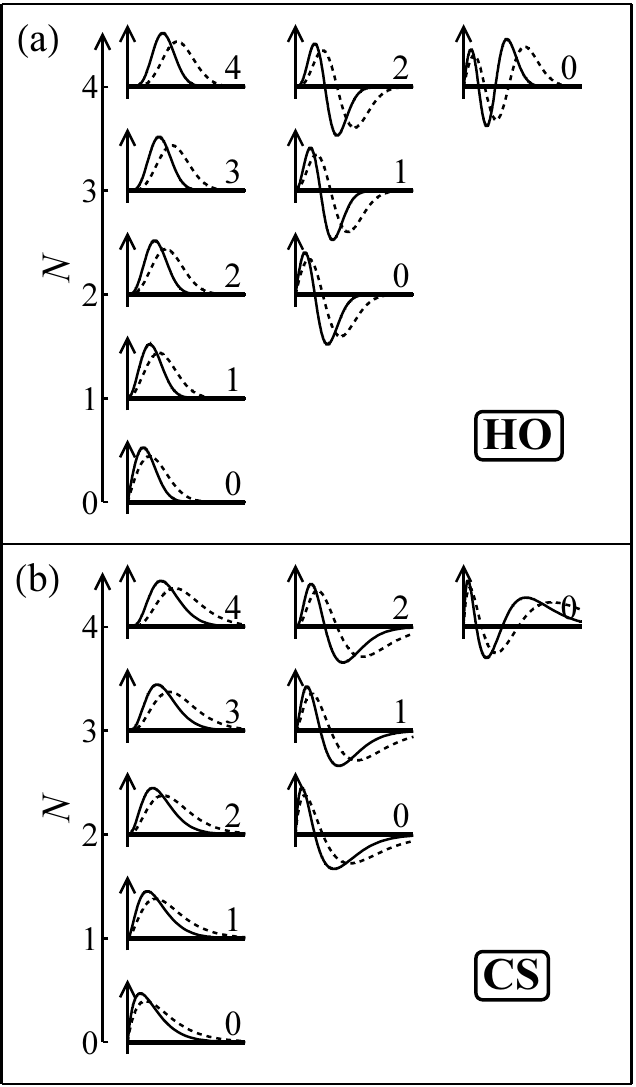}}
\caption{Radial functions (a)~$R_{nl}(b;r)$ of the harmonic oscillator
basis and (b)~$S_{nl}(b_l;r)$ of the Coulomb-Sturmian basis, with
$b_l$ given by the node-matching prescription~(\ref{eqn-bl}).  These
functions are shown arranged according to the harmonic oscillator
principal quantum number
$N\equiv 2n+l$ (see text), and are labeled by $l$.  The dotted curves
show the same functions dilated outward by a factor of
$\sqrt{2}\approx1.414$, corresponding to a factor of $2$ reduction in $\hbar\Omega$.
}
\label{fig-scheme-radial}      
\end{figure}

The harmonic oscillator and Coulomb-Sturmian functions both provide
complete, discrete, orthogonal sets of square-integrable functions, but with Gaussian and exponential asymptotics, respectively.
The oscillator functions~\cite{moshinsky1996:oscillator}, as used in conventional NCCI calculations,
are given by $\wfho_{nlm}(b;\vec{r})=R_{nl}(b;r)Y_{lm}(\uvec{r})/r$,
with radial wave functions
\begin{equation}
\label{eqn-ho-R}
R_{nl}(b;r)\propto(r/b)^{l+1}
L_n^{l+1/2}[(r/b)^2]
e^{-\tfrac12(r/b)^2},
\end{equation}
where  $b$ is the oscillator length.  The Coulomb-Sturmian
functions~\cite{weniger1985:fourier-plane-wave} are given similarly by
$\Lambda_{nlm}(b;\vec{r})=S_{nl}(b;r)Y_{lm}(\uvec{r})/r$,
with radial wave functions 
\begin{equation}
\label{eqn-cs-S}
S_{nl}(b;r)\propto(2r/b)^{l+1}
L_n^{2l+2}(2r/b)
e^{-r/b},
\end{equation}
where $b$ again represents a length scale.  Further discussion may be
found in Ref.~\cite{caprio2012:csbasis}.  In both sets of
functions~(\ref{eqn-ho-R}) and~(\ref{eqn-cs-S}), the $L_n^\alpha$ are
generalized Laguerre polynomials, the $Y_{lm}$ are spherical
harmonics, $n$ is the radial quantum number, and $l$ and $m$ are the
orbital angular momentum and its $z$-projection.
Both sets of radial functions are shown in
Fig.~\ref{fig-scheme-radial}, for comparison.  

For the oscillator functions, the principal quantum number $N\equiv
2n+l$ defines the number of oscillator quanta associated with the
function, or the major shell to which it is assigned, when considered
in the context of an oscillator Hamiltonian with corresponding length
parameter $b$~\cite{suhonen2007:nucleons-nucleus}.  While the
particular combination of $n$ and $l$ represented by $N$ has no
immediate physical significance for the Coulomb-Sturmian functions,
labeling the Coulomb-Sturmian functions by $N$, as in
Fig.~\ref{fig-scheme-radial}(b), can still be of convenience for
consistency with the treatment of the oscillator functions.

For either basis, the nuclear single-particle basis states
$\tket{nljm}$ are defined by coupling of the orbital angular momentum
with the spin, to give total angular momentum $j$.  The many-body
basis is defined by taking antisymmetrized products of these
single-particle states.  

\subsection{Hamiltonian and observables}
\label{sec-methods-hamiltonian}

The structure of the many-body calculation is independent of the
details of the radial basis. The choice of radial basis enters the
many-body calculation only through the values of the Hamiltonian
two-body matrix elements (or higher-body matrix elements, if
higher-body interactions are present), which we must first generate as
the input to the many-body calculation.
The choice of radial basis subsequently also enters into the extraction of
observables (electromagnetic moments and transitions, radii,
\textit{etc.}), from  the densities obtained in the many-body
calculation~\cite{suhonen2007:nucleons-nucleus}.  Here the
relevant inputs are the one-body or two-body matrix elements of the
observable operators with respect to the given basis.

The nuclear Hamiltonian for NCCI calculations has the form $H=\Trel +
V$, where $\Trel$ is the Galilean-invariant, two-body relative kinetic
energy operator, and $V$ is the nucleon-nucleon interaction. A Lawson
term~\cite{gloeckner1974:spurious-com} proportional to the number
$\Ncm$ of center-of-mass oscillator quanta may also be included, to
shift center-of-mass excitations out of the low-lying spectrum.  (The
center-of-mass dynamics for NCCI calculations with the
Coulomb-Sturmian basis, including the effect of a Lawson term, are
investigated in Ref.~\cite{caprio2012:csbasis}.)  However, a Lawson term is not
essential for the present calculations, since we consider only the
ground state, and the calculations of observables (discussed below) make
use only of relative operators, which are, by construction,
insensitive to the center-of-mass degrees of freedom.

The relative kinetic
energy decomposes into one-body and two-body terms as
\begin{equation}
\label{eqn-Trel}
\begin{aligned}
\Trel&\equiv\frac{1}{4Am_N} \sumprime_{ij} (\vec{p}_i-\vec{p}_j)^2\\
&=\frac{1}{2Am_N}\biggl[(A-1) 
\sum_i \vec{p}_i^2
- \sumprime_{ij} \vec{p}_i\cdot\vec{p}_j
\biggr],
\end{aligned}
\end{equation}
where the prime on the summation ${\tsumprime_{ij}}$ over nucleons
indicates $i\neq j$, $A$ is the nuclear mass number, and $m_N$ is the
nucleon mass.  The one-body term may be calculated simply in terms of
one-dimensional radial integrals of the operator $p^2$, with respect
to the radial basis functions.  Since the two-body term is separable,
matrix elements of this term may likewise be calculated in a
straightforward fashion for any radial basis, in terms of radial
integrals of the operators $p$ and angular momentum recoupling
coefficients~\cite{caprio2012:csbasis}.

Calculation of the two-body matrix elements for the interaction,
however, is more involved if one moves to a general radial basis.  The
nucleon-nucleon interaction is defined in relative coordinates.  The
oscillator basis is special, in that matrix elements in a relative
oscillator basis, consisting of functions
$\wfho_{nl}(\vec{r}_1-\vec{r}_2)$, can readily be transformed to the
two-body oscillator basis, consisting of functions
$\wfho_{n_1l_1}(\vec{r}_1)\wfho_{n_2l_2}(\vec{r}_2)$, by the Talmi-Moshinsky
transformation~\cite{moshinsky1996:oscillator}.  We therefore start
from the two-body matrix elements
$\tme{cd;J}{V}{ab;J}$ generated with respect to the oscillator basis, and only
then carry out a change of basis in the
two-body space~\cite{hagen2006:gdm-realistic}.

Specifically, the change of basis for interaction two-body matrix
elements is accomplished by the transformation
\begin{equation}
\label{eqn-tbme-xform}
\tme{\bar{c}\bar{d};J}{V}{\bar{a}\bar{b};J}=
\sum_{abcd} \toverlap{a}{\bar{a}}\toverlap{b}{\bar{b}}
\toverlap{c}{\bar{c}}\toverlap{d}{\bar{d}}
\, \tme{cd;J}{V}{ab;J},
\end{equation}
where we label single-particle orbitals for the oscillator basis by
unbarred symbols $a=(n_al_aj_a)$ and those for the Coulomb-Sturmian
basis by barred symbols $\bar{a}=(\bar n_a
\bar l_a \bar j_a)$.
(See Ref.~\cite{caprio2012:csbasis} for detailed definitions and normalization conventions.)
The coefficients $\toverlap{a}{\bar{a}}$ are obtained from the
one-dimensional overlaps of the harmonic oscillator and
Coulomb-Sturmian radial functions, $\toverlap{R_{nl}}{S_{\bar{n}l}} =
\int_0^\infty dr\, R_{nl}(\bho;r) S_{\bar{n}l}(\bcs;r)$.  It may be
noted that the oscillator length $\bho$~--- with respect to which the
original oscillator two-body matrix elements of the interaction are
represented~--- will in general be different from the length scale $\bcs$
of the Coulomb-Sturmian functions~--- defining the basis for the
many-body calculation.

The change-of-basis transformation
in~(\ref{eqn-tbme-xform}) is, in practice, limited to a finite sum,
\textit{e.g.}, with a shell cutoff $N_a,N_b,N_c,N_d\leq\Ncut$. The
cutoff $\Ncut$ must be chosen high enough to ensure that the results
of the subsequent many-body calculation are cutoff-independent, as
verified by carrying out calculations with differing cutoffs.  The
accuracy obtained for a given cutoff may in general be expected to
depend upon the oscillator and Coulomb-Sturmian length parameters
defining the initial and final bases for the interaction,
respectively, as well as upon the characteristics of the interaction
(\textit{e.g.}, softness), nuclear eigenstates, and observables under
consideration.

The radius observables considered in the study of halo nuclei are the
RMS radii of the point-nucleon distributions: the proton distribution
radius $r_p$, the neutron distribution radius $r_n$, and the combined
matter distribution radius $r_m$.  The RMS radius of the proton, neutron,
and matter distributions are related as
$Ar_m^2=Zr_p^2+Nr_n^2$, and therefore only two out of three of these
may be considered as
independent observables.  Although $r_n$ is perhaps conceptually linked most naturally
to neutron halo structure, 
$r_p$ and $r_m$ are most commonly quoted,
in recognition of experimental considerations (see
Sec.~\ref{sec-results-expt}). 

The radii are all taken
relative to the center of mass of the full set of nucleons,
\textit{i.e.}, protons and neutrons in aggregate, and are obtained
from the expectation values of the relative square-radius operators
defined in~(\ref{eqn-r-defn}).  Much like the $\Trel$ operator
of~(\ref{eqn-Trel}), these are two-body operators which decompose into
one-body and separable two body parts, involving $\sum_i \vec{r}_i^2$
and $\tsumprime_{ij} \vec{r}_i\cdot\vec{r}_j$, respectively, and
evaluation of matrix elements proceeds
similarly~\cite{caprio2012:csbasis}.  Specific relations needed for
evaluating the two-body matrix elements of the proton and neutron
relative square-radius operators with respect to an arbitrary basis
may be found in the Appendix.

\subsection{Basis length parameters and proton-neutron asymmetry}
\label{sec-methods-length}

Any single-particle basis, including~(\ref{eqn-ho-R})
or~(\ref{eqn-cs-S}), has, as a free parameter, an overall length scale,
which we may denote by $b$.  For the oscillator basis, this is
traditionally quoted as the oscillator energy $\hbar\Omega$, where
\begin{equation}
\label{eqn-beta-bho}
b(\hbar\Omega)=\frac{(\hbar c)}{[(m_Nc^2)(\hbar\Omega)]^{1/2}}.
\end{equation}
In deference to the convention of presenting NCCI results as a
function of $\hbar\Omega$ as the basis parameter, we nominally carry over this
relation to define an $\hw$ parameter for general radial bases.  This
$\hbar\Omega$ has no direct physical meaning as an energy scale for
the Coulomb-Sturmian basis.  However, the
inverse square-root dependence remains, so that a factor of two change in $\hbar\Omega$
still describes a factor of $\sqrt{2}$ change in length scale (illustrated for both harmonic oscillator and Coulomb-Sturmian
bases by the dotted curves in Fig.~\ref{fig-scheme-radial}).  

Beyond an overall length scale, there is additional freedom in length
scales which may
be exploited in constructing the basis.  The many-body basis states (antisymmetrized product
states) constructed from a single-particle basis are orthonormal so
long as the single-particle states are orthonormal.  Orthogonality for
single-particle states of different $l$ or $j$ follows entirely from
the angular and spin parts of the wave function.  Only orthogonality
\textit{within} the space of a given $l$ and $j$ follows from the
radial functions, \textit{e.g.}, for the Coulomb-Sturmian functions,
$\toverlap{n'l'j'}{nlj}=\bigl[\int
dr\,S_{n'l}(b;r)\,S_{nl}(b;r)\bigr]\,\delta_{l'l}\delta_{j'j}$.  We
are therefore free to choose $b$ independently, firstly, for each $l$
space (or $j$ space), as $b_l$ (or $b_{lj}$), and, secondly, for
protons and neutrons, as $b_p$ and $b_n$.

The first observation raises the possibility, still to be explored, of
obtaining significant improvements in the efficacy of the basis by
optimizing the $l$-dependence of the length parameter.  For now, we
follow the choice of Ref.~\cite{caprio2012:csbasis} for the
Coulomb-Sturmian functions, which is motivated by more closely
matching the Coulomb-Sturmian functions to the oscillator functions in
the small-$r$ region.  Specifically, $b_l$ is chosen so that the first
node of the $n=1$ Coulomb-Sturmian function for each $l$ aligns with
the first node of the $n=1$ oscillator function for that $l$, which,
from the zeros of the Laguerre polynomials, yields
the prescription
\begin{equation}
\label{eqn-bl}
{b_{l}}=\sqrt{\frac{2}{2l+3}}{b}({{\hbar\Omega}}).
\end{equation}
It is this prescription for $b_l$ which is shown in Fig.~\ref{fig-scheme-radial}(b).

The second observation raises the possibility of proton-neutron
asymmetric length scales, which might be advantageous for nuclei
with significant disparities between the proton and neutron
distributions, in particular, halo nuclei.
Therefore, in the present work, we adopt
\begin{equation}
\label{eqn-blpn}
{b_{l,p}}=\sqrt{\frac{2}{2l+3}}{b}({{\hbar\Omega}}) \qquad {b_{l,n}}={\beta}\sqrt{\frac{2}{2l+3}}{b}({{\hbar\Omega}}),
\end{equation}
where $\beta$ sets an overall relative scale $b_n/b_p$.  For
example, if the solid and dotted curves in
Fig.~\ref{fig-scheme-radial}(b) are taken to represent the proton and
neutron radial functions, respectively, then the figure illustrates
the case in which $\beta\equiv b_n/b_p=\sqrt{2}\approx1.414$.


\section{Calculations for \boldmath$\isotope{He}$ isotopes}
\label{sec-results}

\subsection{Experimental background}
\label{sec-results-expt}

The isotopes $\isotope[6]{He}$ and $\isotope[8]{He}$ are interpreted
as halo nuclei, consisting of a neutron halo surrounding an $\alpha$
core, as reviewed in, \textit{e.g.},
Refs.~\cite{jonson2004:light-dripline,tanihata2013:halo-expt}. The
last neutrons in these isotopes are only weakly bound, with
two-neutron separation energies of $0.97\,\MeV$ and $2.14\,\MeV$,
respectively.  The halo structure is most notably evident in a sudden
increase in the RMS radii of both the proton and matter distributions
along the isotopic chain, summarized in Table~\ref{tab-radii} (see
also Fig.~\ref{fig-crossover-analysis} below).  Moving from
$\isotope[4]{He}$ to $\isotope[6]{He}$, the measured $r_p$ increases
by $\sim32\%$.  This may be understood as resulting from the recoil of
the $\alpha$ core against the halo neutrons~--- \textit{i.e.}, the
presence of the halo neutrons on average displaces the center of mass
of the nucleus away from the center of mass of the $\alpha$
particle~--- as well as possibly receiving a contribution from core
polarization or ``swelling''~\cite{lu2013:laser-neutron-rich}.  An
even greater, though less precisely known, increase in
$r_m$ reflects the extended halo neutron distribution.  The measured
proton and matter radii for $\isotope[8]{He}$ are comparable to those
for $\isotope[6]{He}$.  It is worth briefly summarizing the
experimental situation~--- the origins of the reported radii
and their differences~--- before using them as a baseline for
comparison with the present \textit{ab initio} predictions.

The proton radii $r_p$ are obtained experimentally with comparatively
high precision (better than $0.02\,\fm$).  The charge radius of the
stable isotope $\isotope[4]{He}$ can be measured directly from
electron scattering~\cite{sick2008:escatt-eval}.  The charge radii of
the unstable isotopes $\isotope[6,8]{He}$ are determined indirectly
from isotope shift
data~\cite{wang2004:6he-radius-laser,mueller2007:8he-radius-laser} in
combination with precise mass
measurements~\cite{brodeur2013:6he-8he-mass}.  The RMS radius of the
point-proton distribution is then deduced, after hadronic physics
corrections~\cite{friar1997:charge-radius-correction}, from the
nuclear charge radius. The experimental values for $r_p$ from the
evaluation by Lu \textit{et al.}~\cite{lu2013:laser-neutron-rich} are
$1.462(6)\,\fm$ for $\isotope[4]{He}$, $1.934(9)\,\fm$ for
$\isotope[6]{He}$, and $1.881(17)\,\fm$ for $\isotope[8]{He}$.

The matter radii $r_m$ are obtained with considerably greater
uncertainties, from either nuclear interaction cross
sections~\cite{tanihata1985:radii-he} or proton-nucleus elastic
scattering data~\cite{alkhazov2002:elastic-halo-radii}.  These methods
yield model-dependent and often contradictory results along the
$\isotope{He}$ isotopic chain.

Analyses of the
interaction cross section data for $\isotope[4]{He}$ via the Glauber model yield either
$r_m=1.57(4)\,\fm$~\cite{tanihata1988:radii-be-b-halo} or
$1.63(3)\,\fm$~\cite{tanihata1992:neutron-skins}, depending on
assumptions regarding the parameters for the orbitals
defining the matter distribution.  For $\isotope[4]{He}$, $r_p$ and $r_m$
should be essentially identical by isospin symmetry.  However, these reported
$r_m$ values are substantially larger than and inconsistent, at the stated uncertainties, with the
measured $r_p=1.462(6)\,\fm$.  On the
other hand, elastic scattering yields
$r_m=1.49(3)\,\fm$~\cite{alkhazov2002:elastic-halo-radii}, consistent
with $r_p$.

For $\isotope[6]{He}$, the same Glauber 
analyses of the interaction cross section data yield
$r_m=2.48(4)\,\fm$~\cite{tanihata1988:radii-be-b-halo} or
$2.33(4)\,\fm$~\cite{tanihata1992:neutron-skins}.  However, a few-body
analysis, explicitly considering
$\isotope[6]{He}$ as a correlated system consisting of a core plus two
valence neutrons, suggests a significantly larger value 
$r_m=2.71(4)\,\fm$~\cite{alkhalili2003:inelastic-halo-radii}.  The elastic scattering data yield either
$r_m=2.30(7)\,\fm$ in an analysis assuming Gaussian
asymptotics, or $2.45(10)\,\fm$ in an alternative analysis with
extended (Hankel function)
tails~\cite{alkhazov2002:elastic-halo-radii}.

Finally, for $\isotope[8]{He}$, the Glauber analyses of interaction
cross section data yield
$r_m=2.52(3)\,\fm$~\cite{tanihata1988:radii-be-b-halo} or
$2.49(4)\,\fm$~\cite{tanihata1992:neutron-skins}.  The analyses of
elastic scattering data assuming different asymptotics yield
$r_m=2.45(7)\,\fm$ or
$2.53(8)\,\fm$~\cite{alkhazov2002:elastic-halo-radii}, respectively.

Experimental ranges for $r_m$ encompassing the extreme values
(including uncertainties) of the reported analyses, and identical
to those adopted by Lu
\textit{et al.}~\cite{lu2013:laser-neutron-rich}, are $1.46$--$1.66\,\fm$ for $\isotope[4]{He}$, $2.23$--$2.75\,\fm$ for
$\isotope[6]{He}$, and $2.38$--$2.61\,\fm$ for $\isotope[8]{He}$.  When we compare with theory, 
it is worth bearing in mind that the narrower range of experimental $r_m$ values indicated for
$\isotope[8]{He}$, relative to $\isotope[6]{He}$, does not represent
fundamentally smaller experimental or model uncertainties, but rather
simply a narrower range of attempted model analyses.  The few-body
analysis reported for
$\isotope[6]{He}$~\cite{alkhalili2003:inelastic-halo-radii} is
responsible for raising the upper bound of the experimental range for
this nucleus by $0.2\,\fm$, while no corresponding analysis is
available for $\isotope[8]{He}$.

\subsection{NCCI calculations}
\label{sec-results-calc}

We carry out calculations for the isotopes $\isotope[4,6,8]{He}$ using
both the harmonic oscillator and Coulomb-Sturmian bases.  These
calculations are based on the JISP16 nucleon-nucleon
interaction~\cite{shirokov2007:nn-jisp16}, plus Coulomb interaction.
The bare interaction is used, \textit{i.e.}, without renormalization.
The proton-neutron $M$-scheme code
MFDn~\cite{sternberg2008:ncsm-mfdn-sc08,maris2010:ncsm-mfdn-iccs10,aktulga2013:mfdn-ONLINE}
is employed for the many-body calculations.  

The harmonic oscillator many-body basis is normally truncated
according to the $\Nmax$ scheme, which limits the total number of
oscillator quanta as $\Ntot\equiv\sum_i N_i=\sum_i(2n_i+l_i)\leq
N_0+\Nmax$, where $N_0$ is the minimal number of oscillator quanta for
the given number of protons and neutrons.  We formally carry this
truncation over to the Coulomb-Sturmian basis for the calculations in
the present work, although, as noted in Sec.~\ref{sec-methods-basis},
$N\equiv 2n+l$ no longer has significance as an energy with respect to
a mean field, nor does it lead to the exact factorization of
center-of-mass motion which is obtained with an oscillator basis in
$\Nmax$ truncation (\textit{e.g.}, Ref.~\cite{barrett2013:ncsm}).  Results are calculated with truncations up to
$\Nmax=16$ for $\isotope[4,6]{He}$ and $\Nmax=14$ for
$\isotope[8]{He}$, for both the harmonic oscillator and
Coulomb-Sturmian bases.

\subsection{Results for \boldmath$\isotope[4]{He}$}
\label{sec-results-res-4he}
\begin{figure*}
\centerline{\includegraphics[width=\ifproofpre{0.6}{0.9}\hsize]{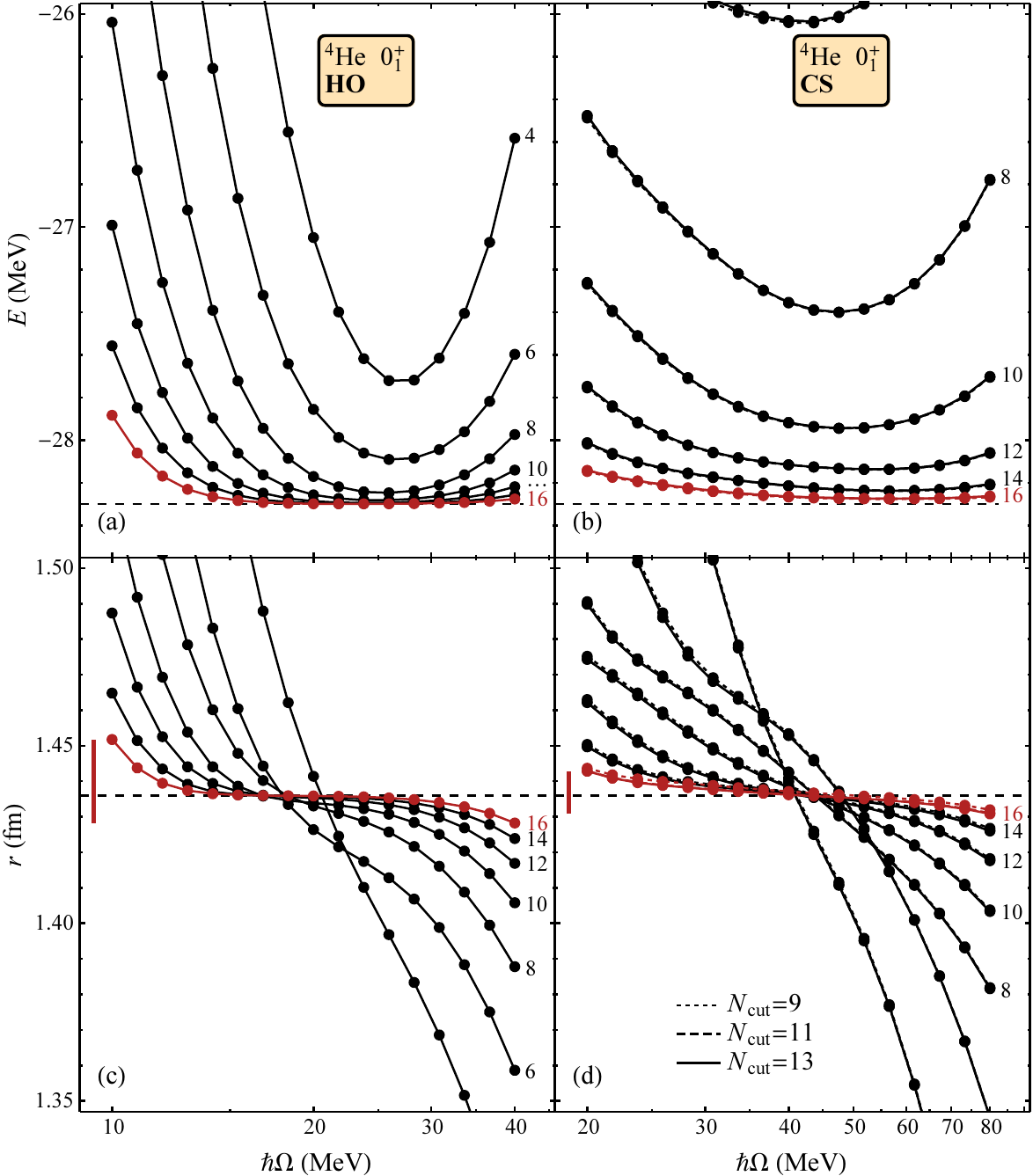}}
\caption{The calculated $\isotope[4]{He}$ ground state energy~(top)
and RMS proton radius $r_p$~(bottom), using the conventional
oscillator~(left) and Coulomb-Sturmian~(right) bases.  These are shown
as functions of the basis $\hbar\Omega$ parameter, for $\Nmax=4$ to
$16$ (as labeled), and for transformation cutoffs $\Ncut=9$, $11$, and
$13$ (Coulomb-Sturmian basis only, indicated by dashing, curves nearly
indistinguishable).  The converged values obtained with the JISP16 interaction
are indicated by dashed horizontal lines.  The spreads in radius
values over this $\hbar\Omega$ range, at the highest $\Nmax$, are
indicated by vertical bars (at bottom).  }
\label{fig-4he-scan}      
\end{figure*}

Let us first consider the calculations for $\isotope[4]{He}$, as the
baseline case.  The computed
ground state energies and proton radii are summarized in
Fig.~\ref{fig-4he-scan}.  Recall that there is no physical meaning in
comparing $\hw$ values directly between oscillator and
Coulomb-Sturmian bases, but that ratios of $\hw$ values within a basis
are  meaningful, serving to indicate the ratio of length scales (Sec.~\ref{sec-methods-length}).
Results are therefore shown consistently over a
factor of four range in $\hw$,
\textit{i.e.}, representing a doubling in basis length scale, for all
bases in the present work, to facilitate comparison across different
bases, and a logarithmic scale is used for $\hw$.

Energy convergence is reached for the harmonic oscillator basis, as
evidenced by approximate $\Nmax$ and $\hw$ independence of the higher
$\Nmax$ results over a range of $\hw$ values, in
Fig.~\ref{fig-4he-scan}(a,b).  Convergence is obtained at the
$\sim10\,\mathrm{keV}$ level by $\Nmax=14$.  The binding energies for
$\isotope[4]{He}$ computed with the Coulomb-Sturmian basis lag
significantly behind those obtained with the oscillator basis, by
about two steps in $\Nmax$.  This should perhaps not be surprising,
given that $\isotope[4]{He}$ is tightly bound, and the structure can
thus be expected to be driven by short-range correlations rather than
asymptotic properties.  

It is important to note that stability with respect to the cutoff in
the change-of-basis transformation~(\ref{eqn-tbme-xform}) has been
obtained~--- calculations with $\Ncut=9$, $11$, and $13$ are virtually
indistinguishable in Fig.~\ref{fig-4he-scan}(b,d).  The transformation
has been carried out from oscillator basis interaction matrix elements
at $\hwint=40\,\MeV$.

Convergence of the computed RMS radii, for both the oscillator and
Coulomb-Sturmian bases, is again indicated by approximate $\Nmax$ and
$\hw$ independence over a range of $\hw$ values, which appears as a
shoulder in the curves of Fig.~\ref{fig-4he-scan}(c,d).  The vertical
bars in Fig.~\ref{fig-4he-scan}(c,d) indicate the spread in radii
obtained (at the highest $\Nmax$) over the range of $\hw$ plotted, to
aid comparison.  The $\hw$ dependence for the Coulomb-Sturmian
calculations appears to be moderately shallower over the range shown,
which spans a factor of four in $\hw$ for each basis.  However, it
should be borne in mind that, since the slopes of the curves in
Fig.~\ref{fig-4he-scan}(c,d) vary significantly with $\hw$, the spread
in radii is sensitive to the particular range of $\hw$ values
chosen, \textit{e.g.}, whether this range is centered on the
variational minimum of the energy calculations or on the crossover
point (it is more simply chosen for purposes of presentation in this
and subsequent figures) and how wide a range is considered.

\subsection{Results for \boldmath$\isotope[6,8]{He}$}
\label{sec-results-res-68he}

\begin{figure*}
\centerline{\includegraphics[width=\ifproofpre{0.6}{0.9}\hsize]{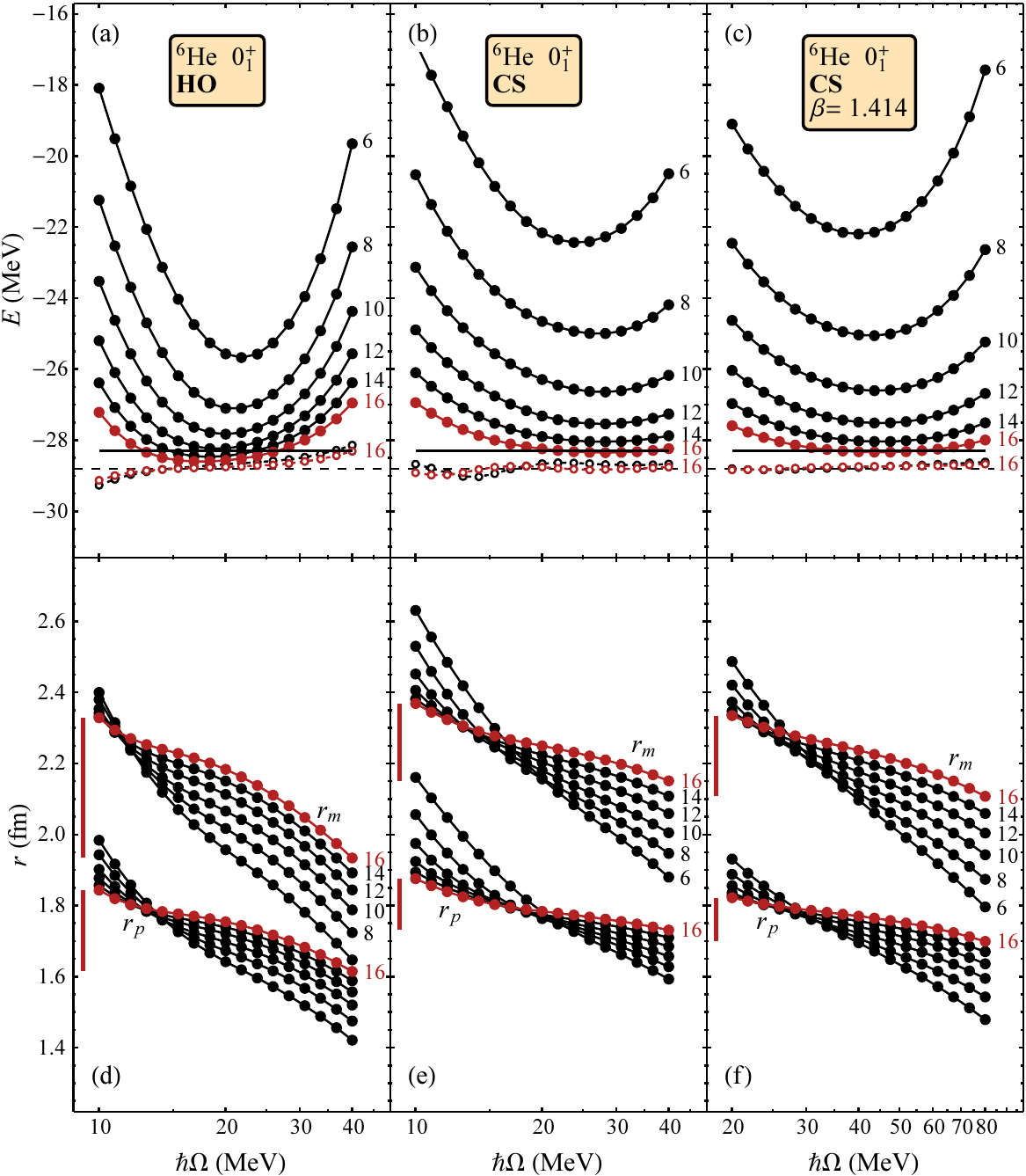}}
\caption{The calculated $\isotope[6]{He}$ ground state energy~(top)
and RMS proton radius $r_p$ and matter radius $r_m$~(bottom), using
the conventional oscillator basis~(left), Coulomb-Sturmian
basis~(center), and proton-neutron asymmetric Coulomb-Sturmian basis
with $\beta=1.414$~(right).  These are shown as functions of the basis
$\hbar\Omega$ parameter, for $\Nmax=6$ to $16$ (as labeled).
Exponentially extrapolated energies from the present calculations are
indicated by open symbols, the best extrapolated energy from
Ref.~\cite{shirokov2014:jisp16-binding} is indicated by the dashed
horizontal line, and the $\isotope[4]{He}+2n$ breakup threshold obtained with
JISP16~\cite{shirokov2014:jisp16-binding} is marked by the solid
horizontal line (at top).  The spreads in radii over this
$\hbar\Omega$ range, at the highest $\Nmax$, are indicated by vertical
bars (at bottom).  
}
\label{fig-6he-scan}      
\end{figure*}
\begin{figure*}
\centerline{\includegraphics[width=\ifproofpre{0.6}{0.9}\hsize]{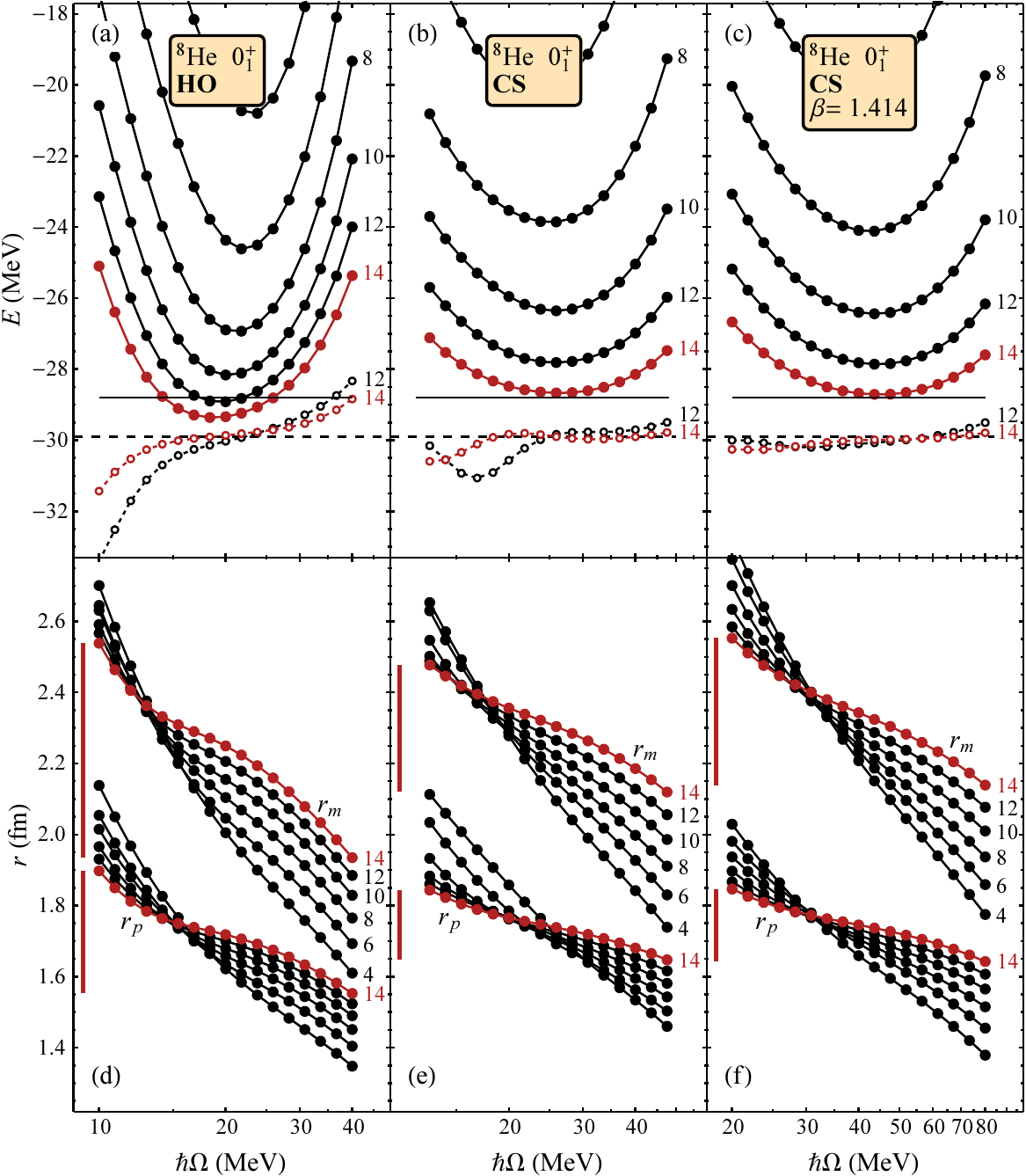}}
\caption{The calculated $\isotope[8]{He}$ ground state energy~(top)
and RMS proton radius $r_p$ and matter radius $r_m$~(bottom), using
the conventional oscillator basis~(left), Coulomb-Sturmian
basis~(center), and proton-neutron asymmetric Coulomb-Sturmian basis
with $\beta=1.414$~(right).  These are shown as functions of the basis
$\hbar\Omega$ parameter, for $\Nmax=4$ to $14$ (as labeled).
Exponentially extrapolated energies from the present calculations are
indicated by open symbols, the best extrapolated energy from
Ref.~\cite{shirokov2014:jisp16-binding} is indicated by the dashed
horizontal line, and the $\isotope[6]{He}+2n$ breakup threshold obtained with
JISP16~\cite{shirokov2014:jisp16-binding} is marked by the solid
horizontal line (at top).  The spreads in radii over this
$\hbar\Omega$ range, at the highest $\Nmax$, are indicated by vertical
bars (at bottom).
}
\label{fig-8he-scan}      
\end{figure*}

Let us now consider the calculations for the halo nuclei $\isotope[6,8]{He}$.  The computed
ground state energies, proton radii, and matter radii are shown in
Figs.~\ref{fig-6he-scan} and~\ref{fig-8he-scan}.  Results are included
(at right in each figure) for a
Coulomb-Sturmian basis with proton-neutron asymmetric length scales
(Sec.~\ref{sec-methods-length}) in
the ratio $\beta\equiv b_n/b_p=1.414$, which is comparable to
the ratio $r_n/r_p$ of neutron and proton distribution radii for these
nuclei.  

Energy convergence in the Coulomb-Sturmian basis lags that of the
harmonic oscillator basis, but less dramatically than seen above for
$\isotope[4]{He}$.  A basic three-point exponential
extrapolation~\cite{maris2009:ncfc} of the energy with respect to
$\Nmax$, at each $\hw$ value, is indicated by the open symbols in
Figs.~\ref{fig-6he-scan}~(top) and~\ref{fig-8he-scan}~(top).  The
extrapolated energy is remarkably $\hw$-independent in the
$\beta=1.414$ calculations, although it should be noted that there is
still some $\Nmax$ dependence as well.  The extrapolated energy
appears to be approximately consistent with the harmonic oscillator
extrapolations.  (The dashed line indicates the best extrapolated
value from harmonic oscillator basis calculations from
Ref.~\cite{shirokov2014:jisp16-binding}, up to $\Nmax=18$ for
$\isotope[6]{He}$ or $\Nmax=14$ for $\isotope[8]{He}$, using a
three-point extrapolation at the $\hbar\Omega$ determined by the
variational energy minimum, yielding binding energies of
$28.803(6)\,\MeV$ and $29.9(2)\,\MeV$ for these isotopes,
respectively.) However, such extrapolations must be viewed with
caution, as both theoretical arguments and empirical studies suggest
that functional forms other than an exponential in $\Nmax$ may be more
appropriate, over at least portions of the $\hw$
range~\cite{coon2012:nscm-ho-regulator,furnstahl2012:ho-extrapolation,more2013:ir-extrapolation}.

Since $\isotope[6]{He}$ and $\isotope[8]{He}$ are weakly bound neutron
halo nuclei, small differences in the calculated binding energy may be
expected to have large effects on the calculated structure, in
particular, whether or not a bound state is even obtained.  While the
JISP16 interaction does bind both $\isotope[6]{He}$ and
$\isotope[8]{He}$ against two-neutron decay, it does so with
two-neutron separation energies of only $0.504(6)\,\MeV$ and
$1.1(2)\,\MeV$, respectively, based on the best extrapolations of
Ref.~\cite{shirokov2014:jisp16-binding}, thus underbinding both nuclei
relative to the experimental values (see Sec.~\ref{sec-results-expt}).
The $2n$ thresholds based on the binding energies obtained with the
JISP16 interaction are indicated in Figs.~\ref{fig-6he-scan} (top)
and~\ref{fig-8he-scan} (top) by the solid horizontal line.  For
$\isotope[6]{He}$, convergence of the energies to the point that the
variational minimum (with respect to $\hbar\Omega$) lies below this
threshold is obtained between $\Nmax=12$ and $14$ for the oscillator
basis calculations [Fig.~\ref{fig-6he-scan}(a)], or between $\Nmax=14$
and $16$ for the Coulomb-Sturmian basis calculations
[Fig.~\ref{fig-6he-scan}(b,c)].  For $\isotope[8]{He}$, the
variational minimum falls below the $2n$ threshold between $\Nmax=10$
and $12$ for the calculations with the oscillator basis
[Fig.~\ref{fig-8he-scan}(a)], while the variational minimum energies
obtained with the Coulomb-Sturmian basis are still just shy of the
threshold for the largest space considered ($\Nmax=14$).  In making
these comparisons, it should be noted that there is little difference
in the variational mimimum energies obtained with the $\beta=1$ or
$\beta=1.414$ calculations, which, \textit{e.g.}, differ by only
$\sim0.01\,\MeV$ for $\isotope[6]{He}$ at $\Nmax=16$
[Fig.~\ref{fig-6he-scan}(b,c)] or $\sim0.04\,\MeV$ for
$\isotope[8]{He}$ at $\Nmax=14$ [Fig.~\ref{fig-8he-scan}(b,c)].  For
both isotopes, the greatest variational gain in binding energy is
actually obtained for an intermediate value for the ratio of proton
and neutron basis length scales, $\beta\approx1.1$--$1.2$ (not shown).

Comparing the results for radii obtained with the different bases, for
$\isotope[6]{He}$ [Fig.~\ref{fig-6he-scan}~(bottom)] and $\isotope[8]{He}$
[Fig.~\ref{fig-8he-scan}~(bottom)], we see that Coulomb-Sturmian calculations (for either $\beta=1$
or $\beta=1.414$) again yield a moderately shallower $\hw$ dependence
than obtained with the harmonic oscillator basis over a wider interval
in $\hbar\Omega$.  On the other hand, the harmonic oscillator basis results give
more of an appearance of localized shouldering.

\section{Radius analysis}
\label{sec-results-crossover}

In examining the dependence of the calculated radii for the He
isotopes (Figs.~\ref{fig-4he-scan}--\ref{fig-8he-scan}) on $\Nmax$ and
$\hbar\Omega$, there is qualitatively similar behavior, across the
bases.  The curves for the radii as functions of $\hbar\Omega$, at
different $\Nmax$, give the appearance of approximately ``converging''
to a common intersection point, at an $\hbar\Omega$ value somewhat
below that of the variational minimum in the energy.  The observation
that, at lower $\hbar\Omega$, the calculated radii decrease with
$\Nmax$, while, at higher $\hbar\Omega$, the calculated radii increase
with $\Nmax$, leaving the calculated radius essentially independent of
$\Nmax$ at the crossover $\hbar\Omega$, might be taken to suggest that
the crossover provides a reasonable estimate of the true converged
radius.

It was therefore proposed in
Refs.~\cite{bogner2008:ncsm-converg-2N,cockrell2012:li-ncfc} that the
radius can be estimated~--- even before convergence is
well-developed~--- by the crossover point.  (Closer inspection reveals
that there is no common intersection point in any strict sense: if we
consider the curves obtained for successive values of $\Nmax$, the
$\hbar\Omega$ value at which these curves cross drifts by several
$\MeV$ as $\Nmax$ increases, generally towards lower $\hbar\Omega$.
Nonetheless, we may consider crossovers between the curves at
successive values of $\Nmax$.)  This is an admittedly
\textit{ad hoc} prescription, rather than a theoretically motivated
extrapolation.  However, while the approach was originally presented
simply in the context of NCCI calculations with the harmonic
oscillator basis, we can now test this approach further and verify
consistency by comparing results obtained from bases with
substantially different underlying single-particle radial functions.

\begin{figure}
\centerline{\includegraphics[width=\ifproofpre{1}{0.6}\hsize]{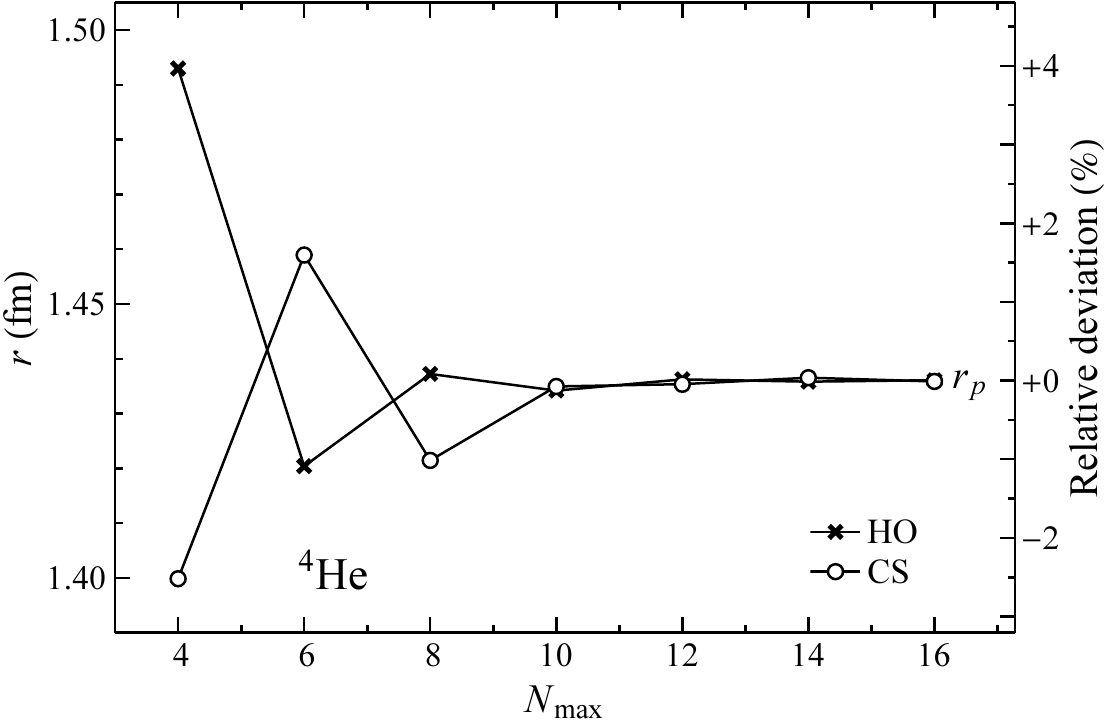}}
\caption{The $\isotope[4]{He}$ ground state RMS proton radius $r_p$,
as estimated from the crossover point (see text), calculated for
the harmonic oscillator and Coulomb-Sturmian bases (as indicated in
the legend).  The relative deviations from the converged value may be
read from the right-hand axis.
}
\label{fig-4he-nmax}      
\end{figure}

We can most directly test the crossover prescription~--- for both
oscillator and Coulomb-Sturmian bases~--- in the case of
$\isotope[4]{He}$, where the final converged value is known.  The
crossover radii are shown as a function of $\Nmax$, for both bases, in
Fig.~\ref{fig-4he-nmax}.  
The curves (of radius as a function of $\hw$
at fixed $\Nmax$) used in deducing these crossovers are computed by
cubic interpolation of the calculated data points at different $\hw$.
The crossovers already serve to estimate the final converged value to
within $\sim0.05\,\fm$ at $\Nmax=6$.\footnote{A similar crossover analysis,
not shown in Fig.~\ref{fig-4he-nmax}, may be carried out for the
matter radius of $\isotope[4]{He}$, yielding marginally smaller values
(by $\sim0.003\,\fm$), since the neutrons are not subject to Coulomb
repulsion.}  The main merit of the approach appears to be that, in the
face of calculated values for the radius which depend smoothly and
strongly on the
basis parameter $\hbar\Omega$, it appears to select out the converged
value more rapidly than, \textit{e.g.}, simply choosing to evaluate the radius at
the $\hbar\Omega$ value which yields the variational minimum in the energy.
\begin{figure}
\centerline{\includegraphics[width=\ifproofpre{1}{0.6}\hsize]{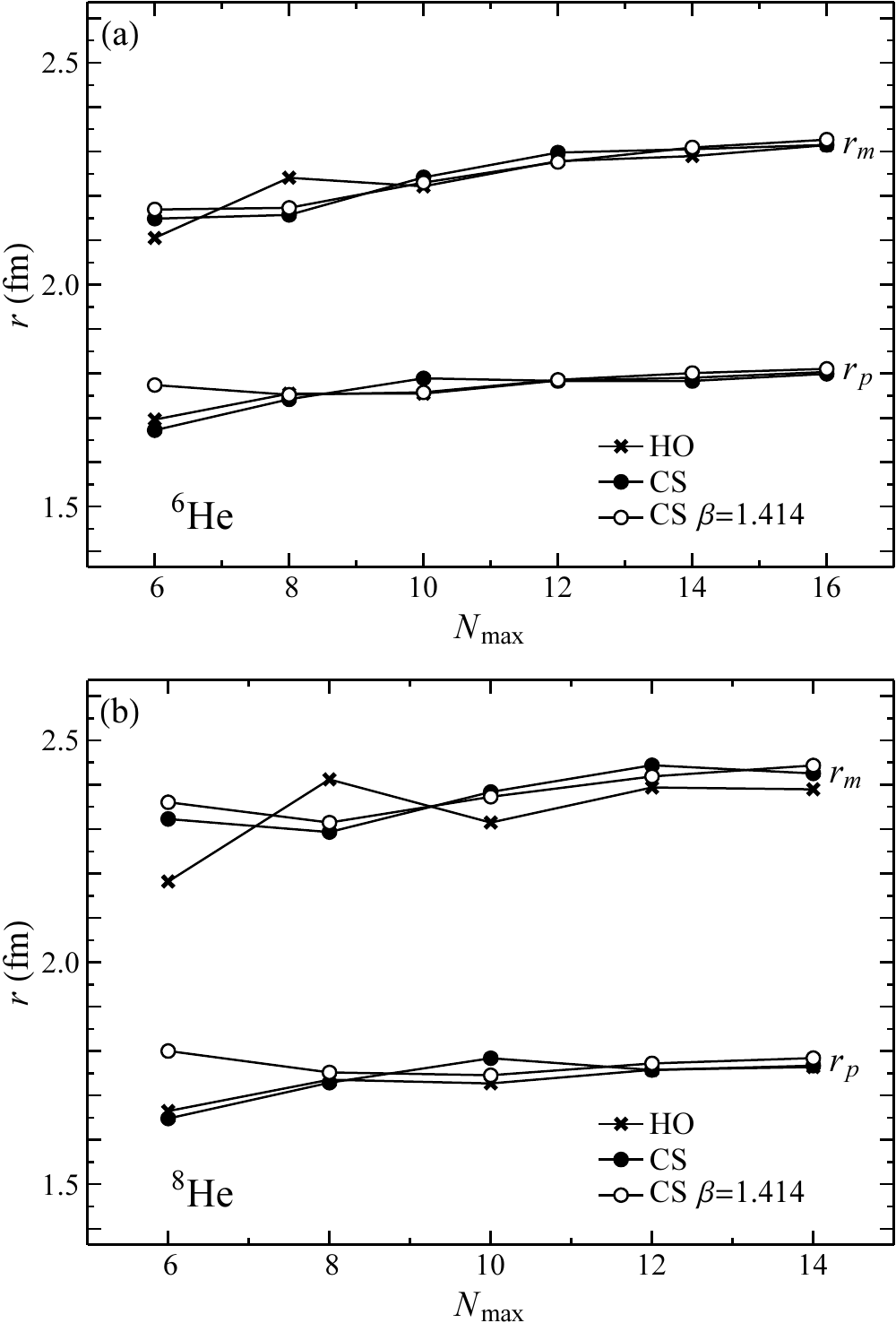}}
\caption{The (a)~$\isotope[6]{He}$ and (b)~$\isotope[8]{He}$ ground state RMS proton radius
$r_p$ (lower curves) and matter radius $r_m$ (upper curves), as
estimated from the crossover point (see text), calculated for the
harmonic oscillator basis, Coulomb-Sturmian basis, and proton-neutron
asymmetric Coulomb-Sturmian basis with $\beta=1.414$ (as indicated in
the legend).  }
\label{fig-68he-nmax}      
\end{figure}
\begin{table}
\caption{The $\isotope[4,6,8]{He}$ proton and matter radii, as estimated
from the crossover point at the highest $\Nmax$ calculated ($\Nmax=16$
for $\isotope[4,6]{He}$ and $\Nmax=14$ for $\isotope[8]{He}$), for the
harmonic oscillator basis (HO), Coulomb-Sturmian basis (CS), and
proton-neutron asymmetric Coulomb-Sturmian basis with $\beta=1.414$
(CS $\beta$).  Experimental values or ranges (see
Sec.~\ref{sec-results-expt}) are also given.  All radii are in
$\fm$. }
\label{tab-radii}
\begin{center}
\begin{ruledtabular}
\begin{tabular}{llD{.}{.}{1.6}D{.}{.}{1.5}D{.}{.}{1.5}}
& & \multicolumn{1}{c}{$\isotope[4]{He}$} & \multicolumn{1}{c}{$\isotope[6]{He}$} & \multicolumn{1}{c}{$\isotope[8]{He}$} 
\\
\hline
$r_p$ 
& HO & 1.4361 & 1.803 & 1.764 
\\
& CS & 1.4358 & 1.799 & 1.767
\\
& CS $\beta$ & \multicolumn{1}{c}{---} & 1.810 & 1.784
\\
\cline{2-5}
& Expt. & 1.462(6) & 1.934(9) & 1.881(17)
\\
\hline
$r_m$ 
& HO & 1.4335 & 2.314 & 2.390
\\
& CS & 1.4332 & 2.315 & 2.425
\\
& CS $\beta$ & \multicolumn{1}{c}{---} & 2.327 & 2.443
\\
\cline{2-5}
& Expt. & \multicolumn{1}{c}{$1.46$--$1.66$} &
\multicolumn{1}{c}{$2.23$--$2.75$} &  \multicolumn{1}{c}{$2.38$--$2.61$}
\end{tabular}
\end{ruledtabular}
\raggedright
\end{center}
\end{table}
\begin{figure}
\centerline{\includegraphics[width=\ifproofpre{1}{0.6}\hsize]{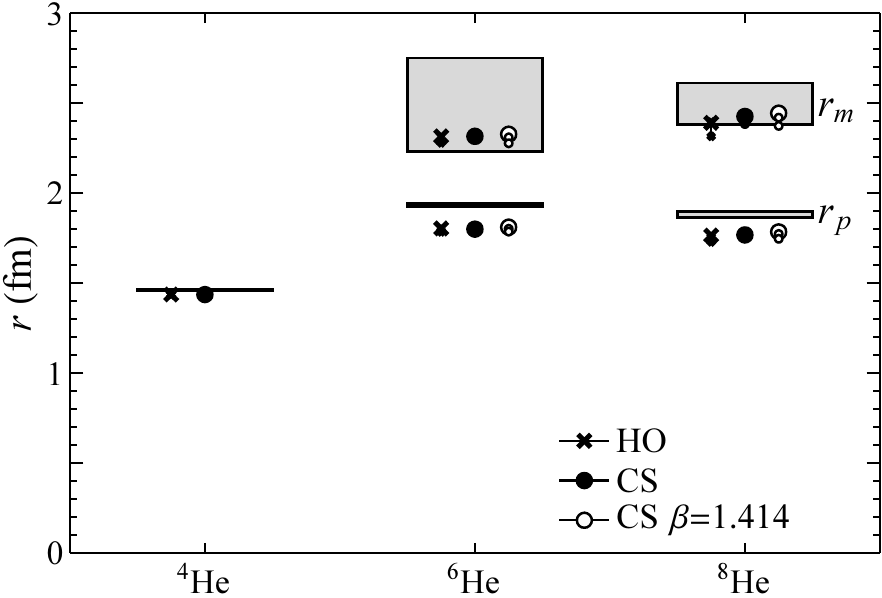}}
\caption{The $\isotope[4]{He}$ proton radius
and $\isotope[6,8]{He}$ proton and matter radii, as estimated from the
crossover point, for the harmonic oscillator basis, Coulomb-Sturmian
basis, and proton-neutron asymmetric Coulomb-Sturmian basis with
$\beta=1.414$.  For each of these bases, the extracted radii
are shown for the highest three successive $\Nmax$ values
($12\leq\Nmax\leq16$ for $\isotope[4,6]{He}$ or $10\leq\Nmax\leq14$
for $\isotope[8]{He}$), in some cases visually indistinguishable, with
the largest symbol indicating the highest $\Nmax$ value.  Experimental
values or ranges are shown as horizontal bands.  }
\label{fig-crossover-analysis}      
\end{figure}

The extracted crossover radii for $\isotope[6,8]{He}$ are shown, as
functions of $\Nmax$, in Fig.~\ref{fig-68he-nmax}.  The radii obtained
for the Coulomb-Sturmian calculations with different ratios of neutron
and proton length scales ($\beta=1$ and $1.414$) track each other
closely from $\Nmax\approx8$ onward, agreeing with each other to
within $\sim0.1\,\fm$.  For $r_p$, the values are stable with respect
to $\Nmax$ and agree with the values obtained from the harmonic
oscillator basis crossover as well.  For $r_m$, it appears that the
values might be drifting systematically upward with $\Nmax$, although
they do remain within an $\sim0.2\,\fm$ range from $\Nmax=8$ to the
highest $\Nmax$ calculated.  Therefore, although we can extract a
result based on this highest $\Nmax$ (as discussed below), it is not possible to
give a definitive value for $r_m$.

An overview of the predicted evolution of the radius observables along
the $\isotope{He}$ isotopic chain, and a comparison with experimental
values, is provided in Fig.~\ref{fig-crossover-analysis}.  (Although
the dependence of the extracted crossover radius on $\Nmax$ was shown
in Figs.~\ref{fig-4he-nmax} and~\ref{fig-68he-nmax}, here it is
helpful to directly see the stability of each radius with respect to
$\Nmax$, by overlaying the results obtained for the three highest
successive $\Nmax$ values, with the largest symbol indicating the
result for the highest $\Nmax$ value.)  The radii obtained at the
highest $\Nmax$, for each basis, are summarized in
Table~\ref{tab-radii}. For each radius considered, the values obtained
from the calculations with different bases are consistent to within
$\sim0.02\,\fm$, or $\sim0.05\,\fm$ in the case of the
$\isotope[6]{He}$ matter radius.\footnote{The present values for the
$\isotope[6,8]{He}$ radii are consistent with
estimates~\cite{constantinou-IP} obtained, from the same NCCI
calculations, by infrared oscillator basis extrapolation methods of
the type proposed in
Refs.~\cite{furnstahl2012:ho-extrapolation,more2013:ir-extrapolation,furnstahl2014:ir-expansion}.
(The detailed results are sensitive to the range of $\Nmax$ and $\hw$
values included in the extrapolation procedure, as well as to the
prescription used for the infrared cutoff
parameter~\cite{furnstahl-PREPRINT:ir-extrapolation-cc-16o}.)  The
present values are also consistent with estimates $r_p=1.84(8)\,\fm$
and $r_m=2.43(19)\,\fm$ obtained from calculations for $\isotope[6]{He}$ using Woods-Saxon bases, under the
JISP16 interaction, in
Refs.~\cite{negoita2010:diss,negoita-IP}.}  The radii for these light
nuclei are also accessible to other \textit{ab initio} methods~---
results have recently been reported based on the effective interaction
hyperspherical harmonic (EIHH)
method~\cite{bacca2012:6he-hyperspherical} and Green's function Monte
Carlo (GFMC) method~\cite{lu2013:laser-neutron-rich} and could be
extracted from calculations based upon the no-core shell
model/resonating group method
(NCSM/RGM)~\cite{quaglioni2013:ncsm-rgm-cluster-6he}~--- suggesting
the possibility of benchmarking calculations carried out for the same
interaction under different calculational approaches and extrapolation
schemes~\cite{coon2013:ncsm-extrapolation-ntse13-DUMMY,furnstahl2014:ir-expansion}.

The proton radius calculated for
$\isotope[4]{He}$ matches the experimental result to within
$\sim0.02\,\fm$.  Indeed, this is perhaps an unreasonably good level
of agreement to expect from imperfectly known internucleon
interactions.  In any case, it is at the same scale as
systematic uncertainties in the experimental corrections for the
proton size from hadronic physics~\cite{lu2013:laser-neutron-rich}.

The present calculations with the JISP16 interaction qualitatively
reproduce the observed jump in radii.  From $\isotope[4]{He}$ to
$\isotope[6]{He}$, the calculated $r_p$ increases by $25\%$~---
quantitatively somewhat short of the measured $32\%$ increase~--- then
remains essentially unchanged for $\isotope[8]{He}$.  The calculated
$r_m$ increases by $62\%$ from $\isotope[4]{He}$ to $\isotope[6]{He}$,
again remaining essentially unchanged for $\isotope[8]{He}$.  These
matter radii are in good agreement with the elastic scattering
measurements (Sec.~\ref{sec-results-expt}), \textit{i.e.}, with the
lower end of the experimental range.

\section{Conclusion}
\label{sec-concl}

The present work is, in various respects, an investigation of
computational methods (alternative radial bases for the NCCI
approach), an investigation of analysis methods (for extracting an
estimator of the converged radius from still-unconverged
calculations), and an investigation of a physical problem (\textit{ab
initio} prediction of halo structure in the $\isotope{He}$ isotopes).

From the computational viewpoint, the NCCI approach has been applied
with bases incorporating realistic exponential asymptotics (the
Coulomb-Sturmian functions) and proton-neutron asymmetry (in
recognition of the physical asymmetry of the system).  Calculations
with the Coulomb-Sturmian basis are found to be valuable in predicting
RMS radius observables for the $\isotope{He}$ isotopes subject to the
JISP16 interaction.  Convergence of the binding energy is moderately
slower than with the harmonic oscillator basis.  This appears to be at
least partially offset by more stable extrapolation properties when
the basic exponential extrapolation scheme is used.  Calculations of
the RMS radii of $\isotope[6,8]{He}$ appear to show improved $\hw$
independence with the Coulomb-Sturmian basis.  However, for both
observables, more complete, theoretically motivated extrapolation
studies are needed.

It would seem that a principal underlying challenge to devising an
appropriate expansion basis is the compromise involved in addressing
both the core nucleons and the halo nucleons with basis functions
sharing the same length parameter, and thus the same rate of
exponential fall-off in the asymptotic region.  A single-particle
basis encompassing functions with differing length scales (as
encountered in atomic electron-structure
calculations~\cite{helgaker2000:electron-structure}) might be expected
to provide greater efficiency in describing halo structure.
Further optimization of the Coulomb-Sturmian basis is also likely possible through variation
of the $l$ dependence of the length parameters, potentially yielding
improved convergence (analogous optimizations are again important for rapid
convergence in electron-structure
calculations).

In the present work, perhaps the most direct benefit of going beyond
the oscillator basis lies simply in being able to compare calculations
obtained with qualitatively different basis sets, and thereby to
verify the robustness of the estimated observable values extracted
from still-unconverged calculations.  The crossover prescription for
radius observables, although
\textit{ad hoc}, is found to yield consistent results, to within $\sim0.02$--$0.05\,\fm$, between bases
with substantially different underlying single-particle radial
functions.

These results give estimates for the proton and matter radii of the
$\isotope{He}$ halo nuclei, based on \textit{ab initio} calculations,
with the JISP16 interaction.  The distinctive trend in radii
along the $\isotope{He}$ isotopic chain, indicative of the onset of
halo structure, is qualitatively reproduced.  More quantitatively, the
proton radii of the halo isotopes are underestimated, relative to
experiment, while the calculated matter radii favor the lower end of
the experimental range.


\begin{acknowledgments}
We thank S.~Quaglioni, S.~Bacca, M.~Brodeur, J.~Parkhill, and A.~M.~Shirokov for valuable
discussions and A.~E.~McCoy and Ch.~Constantinou for comments on the manuscript.  This work
was supported by the Research Corporation for Science Advancement
through the Cottrell Scholar program, by the US Department of Energy
under Grants No.~DE-FG02-95ER-40934, DESC0008485 (SciDAC/NUCLEI), and
DE-FG02-87ER40371, and by the US National Science Foundation under
Grant No.~0904782. Computational resources were provided by the
National Energy Research Scientific Computing Center (NERSC), which is
supported by the Office of Science of the U.S. Department of Energy
under Contract No.~DE-AC02-05CH11231.
\end{acknowledgments}


\appendix

\section{Proton and neutron squared-radius operators}
\label{app-angular}

The RMS radii of the point-nucleon proton, neutron, or matter
distributions, relative to the center of
mass (Sec.~\ref{sec-methods}), are calculated as
$r_p\equiv\tbracket{r_p^2}^{1/2}$, $r_n\equiv\tbracket{r_n^2}^{1/2}$,
or $r_m\equiv\tbracket{\rrel^2}^{1/2}$, in terms of relative mean squared radius
operators, which are defined by~\cite{bacca2012:6he-hyperspherical}
\begin{equation}
\label{eqn-r-defn}
\begin{aligned}
r_p^2&=\frac{1}{N_p}\sum_i \delta_{p,i} (\vec{r}_i-\vec{R})^2\\
r_n^2&=\frac{1}{N_n}\sum_i \delta_{n,i} (\vec{r}_i-\vec{R})^2\\
\rrel^2&=\frac{1}{A}\sum_i(\vec{r}_i-\vec{R})^2,
\end{aligned}
\end{equation}
in terms of the
center-of-mass position operator
\begin{equation}
\label{eqn-R-defn}
\vec{R}=\frac{1}{A}\sum_i\vec{r}_i,
\end{equation}
where we define the shorthands $\delta_{p,i}=\tfrac12(1+\tau_{z,i})$
and $\delta_{n,i}=\tfrac12(1-\tau_{z,i})$ (with $\tau_{z}=+1$ for
protons and $-1$ for neutrons) to
select proton and neutron indices, respectively, and we
denote the proton and neutron numbers by $N_p\,(\equiv Z)$ and
$N_n\,(\equiv N)$ to provide greater uniformity between the
expressions for the proton and neutron radii below.
The operators in~(\ref{eqn-r-defn}) are two-body operators, due to the
subtraction of the center-of-mass coordinate.  Thus, in order to calculate their expectation values
within a many-body wave function, the two-body matrix
elements of these operators are required, with respect to the basis for the
calculation.  
In this appendix, we summarize certain operator relations needed for
evaluating the two-body matrix elements of these operators.

The $\rrel^2$ operator, as defined in~(\ref{eqn-r-defn}), can be
reexpressed in forms more suitable for
evaluation of two-body matrix elements, as outlined in
Appendix~A of Ref.~\cite{caprio2012:csbasis}.  On the one hand,
$\rrel^2$ can
be expressed in the standard form for a two-body operator,
\textit{i.e.}, as a double sum
over distinct particle indices, as
\begin{equation}
\label{eqn-rrel-2b}
\rrel^2=\frac{1}{2A^2}\sumprime_{ij}(\vec{r}_i-\vec{r}_j)^2,
\end{equation}
which we will use below.
On the other hand, expanding the square in~(\ref{eqn-r-defn}) or~(\ref{eqn-rrel-2b})  gives an alternate expression for
$\rrel^2$ in
terms of one-body and separable two-body parts
\begin{equation}
\label{eqn-rrel-expanded}
\rrel^2=\frac{(A-1)}{A^2}\sum_{i}r_i^2-\frac{1}{A^2}\sumprime_{ij}\vec{r}_i\cdot\vec{r}_j.
\end{equation}
This latter expression may be used to evaluate the two-body matrix
elements $\tme{cd;J}{\rrel^2}{ab;J}$ in a
straightforward fashion, from the
radial integrals of the $r$ and $r^2$ operators with respect to the
given single-particle basis, as elaborated in
Sec.~III\,D of Ref.~\cite{caprio2012:csbasis}.

The two-body matrix elements of the operators $r_p^2$ and $r_n^2$ may,
conveniently, be deduced from those already obtained for the operator
$\rrel^2$.  To establish the relationship, it is
helpful to first define the restrictions of $\rrel^2$ to the
proton-proton, proton-neutron, and neutron-neutron sectors
\begin{equation}
\label{eqn-resticted}
\begin{aligned}
\rrelpp^2&=\frac{1}{2A^2}\sumprime_{ij} \delta_{p,i}\delta_{p,j} (\vec{r}_i-\vec{r}_j)^2.\\
\rrelpn^2&=\frac{1}{2A^2}\sumprime_{ij} (\delta_{p,i}\delta_{n,j}+\delta_{n,i}\delta_{p,j}) (\vec{r}_i-\vec{r}_j)^2.\\
\rrelnn^2&=\frac{1}{2A^2}\sumprime_{ij} \delta_{n,i}\delta_{n,j} (\vec{r}_i-\vec{r}_j)^2.
\end{aligned}
\end{equation}
Thus, 
\begin{equation}
\label{eqn-rrel-rrelxx}
\rrel^2=\rrelpp^2+\rrelpn^2+\rrelnn^2.
\end{equation}

These operators are convenient to consider in the evaluation of
two-body matrix elements, since their matrix elements are simply
connected to those of $\rrel^2$.  The matrix elements of
$\rrelpp^2$ are obtained by restricting those of $\rrel^2$ to the
proton-proton sector, \textit{i.e.},
$\tme{cd;J}{\rrelpp^2}{ab;J}_{pp}=\tme{cd;J}{\rrel^2}{ab;J}_{pp}$,
with matrix elements in other sectors vanishing. Similarly,
the matrix elements of
$\rrelpn^2$ are obtained by restricting those of $\rrel^2$ to the
proton-neutron sector, and the matrix elements of
$\rrelnn^2$ are obtained by restricting those of $\rrel^2$ to the
neutron-neutron sector.  
Then, the relative square-radius operators of interest are expressed
in terms of these as
\begin{equation}
\label{eqn-rp-rrelxx}
r_p^2=\frac{(2A-N_p)}{N_p}\rrelpp^2+\frac{(A-N_p)}{N_p}\rrelpn^2-\rrelnn^2,
\end{equation}
and, interchanging labels ($p\leftrightarrow n$),
\begin{equation}
\label{eqn-rn-rrelxx}
r_n^2=-\rrelpp^2+\frac{(A-N_n)}{N_n}\rrelpn^2+\frac{(2A-N_n)}{N_n}\rrelnn^2.
\end{equation}
The equivalence of the expressions for $r_p^2$ and $r_n^2$ in~(\ref{eqn-r-defn})
to those in~(\ref{eqn-rp-rrelxx}) and~(\ref{eqn-rn-rrelxx}) may be
verified in a straightforward fashion [\textit{e.g.}, by expanding the squares in both expressions, so
that the resulting expressions contain only one-body and separable
two-body terms as in~(\ref{eqn-rrel-expanded}), and comparing
terms].  The relations~(\ref{eqn-rp-rrelxx}) and~(\ref{eqn-rn-rrelxx})
immediately allow the two-body matrix elements of $r_p^2$ and $r_n^2$
to be obtained in terms of those of $\rrel^2$.  For instance, from the first term
of~(\ref{eqn-rp-rrelxx}), we read off
$\tme{cd;J}{r_p^2}{ab;J}_{pp}=[(2A-N_p)/N_p]\tme{cd;J}{\rrel^2}{ab;J}_{pp}$.

\vfill


\providecommand{\APSLONG}{}
\providecommand{\ELSEVIER}{}



\clearpage


\begin{thebibliography}{54}
\expandafter\ifx\csname natexlab\endcsname\relax\def\natexlab#1{#1}\fi
\expandafter\ifx\csname bibnamefont\endcsname\relax
  \def\bibnamefont#1{#1}\fi
\expandafter\ifx\csname bibfnamefont\endcsname\relax
  \def\bibfnamefont#1{#1}\fi
\expandafter\ifx\csname citenamefont\endcsname\relax
  \def\citenamefont#1{#1}\fi
\expandafter\ifx\csname url\endcsname\relax
  \def\url#1{\texttt{#1}}\fi
\expandafter\ifx\csname urlprefix\endcsname\relax\def\urlprefix{URL }\fi
\providecommand{\bibinfo}[2]{#2}
\providecommand{\eprint}[2][arXiv]{\url{#1:#2}}
\renewcommand{\eprint}[2][arXiv]{\url{#1:#2}}

\bibitem{navratil2000:12c-ab-initio}
\bibinfo{author}{\bibfnamefont{P.}~\bibnamefont{Navr\'{a}til}},
  \bibinfo{author}{\bibfnamefont{J.~P.} \bibnamefont{Vary}}, \bibnamefont{and}
  \bibinfo{author}{\bibfnamefont{B.~R.} \bibnamefont{Barrett}},
  \bibinfo{journal}{Phys. Rev. Lett.} \textbf{\bibinfo{volume}{84}},
  \bibinfo{pages}{5728} (\bibinfo{year}{2000}).

\bibitem{navratil2000:12c-ncsm}
\bibinfo{author}{\bibfnamefont{P.}~\bibnamefont{Navr\'{a}til}},
  \bibinfo{author}{\bibfnamefont{J.~P.} \bibnamefont{Vary}}, \bibnamefont{and}
  \bibinfo{author}{\bibfnamefont{B.~R.} \bibnamefont{Barrett}},
  \bibinfo{journal}{Phys. Rev. C} \textbf{\bibinfo{volume}{62}},
  \bibinfo{pages}{054311} (\bibinfo{year}{2000}).

\bibitem{barrett2013:ncsm}
\bibinfo{author}{\bibfnamefont{B.~R.} \bibnamefont{Barrett}},
  \bibinfo{author}{\bibfnamefont{P.}~\bibnamefont{Navr\'{a}til}},
  \bibnamefont{and} \bibinfo{author}{\bibfnamefont{J.~P.} \bibnamefont{Vary}},
  \bibinfo{journal}{Prog. Part. Nucl. Phys.} \textbf{\bibinfo{volume}{69}},
  \bibinfo{pages}{131} (\bibinfo{year}{2013}).

\bibitem{forssen2008:ncsm-sequences}
\bibinfo{author}{\bibfnamefont{C.}~\bibnamefont{Forssen}},
  \bibinfo{author}{\bibfnamefont{J.~P.} \bibnamefont{Vary}},
  \bibinfo{author}{\bibfnamefont{E.}~\bibnamefont{Caurier}}, \bibnamefont{and}
  \bibinfo{author}{\bibfnamefont{P.}~\bibnamefont{Navratil}},
  \bibinfo{journal}{Phys. Rev. C} \textbf{\bibinfo{volume}{77}},
  \bibinfo{pages}{024301} (\bibinfo{year}{2008}).

\bibitem{bogner2008:ncsm-converg-2N}
\bibinfo{author}{\bibfnamefont{S.~K.} \bibnamefont{Bogner}},
  \bibinfo{author}{\bibfnamefont{R.~J.} \bibnamefont{Furnstahl}},
  \bibinfo{author}{\bibfnamefont{P.}~\bibnamefont{Maris}},
  \bibinfo{author}{\bibfnamefont{R.~J.} \bibnamefont{Perry}},
  \bibinfo{author}{\bibfnamefont{A.}~\bibnamefont{Schwenk}}, \bibnamefont{and}
  \bibinfo{author}{\bibfnamefont{J.}~\bibnamefont{Vary}},
  \bibinfo{journal}{Nucl. Phys. A} \textbf{\bibinfo{volume}{801}},
  \bibinfo{pages}{21} (\bibinfo{year}{2008}).

\bibitem{maris2009:ncfc}
\bibinfo{author}{\bibfnamefont{P.}~\bibnamefont{Maris}},
  \bibinfo{author}{\bibfnamefont{J.~P.} \bibnamefont{Vary}}, \bibnamefont{and}
  \bibinfo{author}{\bibfnamefont{A.~M.} \bibnamefont{Shirokov}},
  \bibinfo{journal}{Phys. Rev. C} \textbf{\bibinfo{volume}{79}},
  \bibinfo{pages}{014308} (\bibinfo{year}{2009}).

\bibitem{coon2012:nscm-ho-regulator}
\bibinfo{author}{\bibfnamefont{S.~A.} \bibnamefont{Coon}},
  \bibinfo{author}{\bibfnamefont{M.~I.} \bibnamefont{Avetian}},
  \bibinfo{author}{\bibfnamefont{M.~K.~G.} \bibnamefont{Kruse}},
  \bibinfo{author}{\bibfnamefont{U.}~\bibnamefont{van Kolck}},
  \bibinfo{author}{\bibfnamefont{P.}~\bibnamefont{Maris}}, \bibnamefont{and}
  \bibinfo{author}{\bibfnamefont{J.~P.} \bibnamefont{Vary}},
  \bibinfo{journal}{Phys. Rev. C} \textbf{\bibinfo{volume}{86}},
  \bibinfo{pages}{054002} (\bibinfo{year}{2012}).

\bibitem{furnstahl2012:ho-extrapolation}
\bibinfo{author}{\bibfnamefont{R.~J.} \bibnamefont{Furnstahl}},
  \bibinfo{author}{\bibfnamefont{G.}~\bibnamefont{Hagen}}, \bibnamefont{and}
  \bibinfo{author}{\bibfnamefont{T.}~\bibnamefont{Papenbrock}},
  \bibinfo{journal}{Phys. Rev. C} \textbf{\bibinfo{volume}{86}},
  \bibinfo{pages}{031301} (\bibinfo{year}{2012}).

\bibitem{more2013:ir-extrapolation}
\bibinfo{author}{\bibfnamefont{S.~N.} \bibnamefont{More}},
  \bibinfo{author}{\bibfnamefont{A.}~\bibnamefont{Ekstrom}},
  \bibinfo{author}{\bibfnamefont{R.~J.} \bibnamefont{Furnstahl}},
  \bibinfo{author}{\bibfnamefont{G.}~\bibnamefont{Hagen}}, \bibnamefont{and}
  \bibinfo{author}{\bibfnamefont{T.}~\bibnamefont{Papenbrock}},
  \bibinfo{journal}{Phys. Rev. C} \textbf{\bibinfo{volume}{87}},
  \bibinfo{pages}{044326} (\bibinfo{year}{2013}).

\bibitem{jurgenson2013:ncsm-srg-pshell}
\bibinfo{author}{\bibfnamefont{E.~D.} \bibnamefont{Jurgenson}},
  \bibinfo{author}{\bibfnamefont{P.}~\bibnamefont{Maris}},
  \bibinfo{author}{\bibfnamefont{R.~J.} \bibnamefont{Furnstahl}},
  \bibinfo{author}{\bibfnamefont{P.}~\bibnamefont{Navratil}},
  \bibinfo{author}{\bibfnamefont{W.~E.} \bibnamefont{Ormand}},
  \bibnamefont{and} \bibinfo{author}{\bibfnamefont{J.~P.} \bibnamefont{Vary}},
  \bibinfo{journal}{Phys. Rev. C} \textbf{\bibinfo{volume}{87}},
  \bibinfo{pages}{054312} (\bibinfo{year}{2013}).

\bibitem{jonson2004:light-dripline}
\bibinfo{author}{\bibfnamefont{B.}~\bibnamefont{Jonson}},
  \bibinfo{journal}{Phys. Rep.} \textbf{\bibinfo{volume}{389}},
  \bibinfo{pages}{1} (\bibinfo{year}{2004}).

\bibitem{tanihata2013:halo-expt}
\bibinfo{author}{\bibfnamefont{I.}~\bibnamefont{Tanihata}},
  \bibinfo{author}{\bibfnamefont{H.}~\bibnamefont{Savajols}}, \bibnamefont{and}
  \bibinfo{author}{\bibfnamefont{R.}~\bibnamefont{Kanungo}},
  \bibinfo{journal}{Prog. Part. Nucl. Phys.} \textbf{\bibinfo{volume}{68}},
  \bibinfo{pages}{215} (\bibinfo{year}{2013}).

\bibitem{quaglioni2009:ncsm-rgm}
\bibinfo{author}{\bibfnamefont{S.}~\bibnamefont{Quaglioni}} \bibnamefont{and}
  \bibinfo{author}{\bibfnamefont{P.}~\bibnamefont{Navr{\'a}til}},
  \bibinfo{journal}{Phys. Rev. C} \textbf{\bibinfo{volume}{79}},
  \bibinfo{pages}{044606} (\bibinfo{year}{2009}).

\bibitem{cockrell2012:li-ncfc}
\bibinfo{author}{\bibfnamefont{C.}~\bibnamefont{Cockrell}},
  \bibinfo{author}{\bibfnamefont{J.~P.} \bibnamefont{Vary}}, \bibnamefont{and}
  \bibinfo{author}{\bibfnamefont{P.}~\bibnamefont{Maris}},
  \bibinfo{journal}{Phys. Rev. C} \textbf{\bibinfo{volume}{86}},
  \bibinfo{pages}{034325} (\bibinfo{year}{2012}).

\bibitem{maris2013:ncsm-pshell}
\bibinfo{author}{\bibfnamefont{P.}~\bibnamefont{Maris}} \bibnamefont{and}
  \bibinfo{author}{\bibfnamefont{J.~P.} \bibnamefont{Vary}},
  \bibinfo{journal}{Int. J. Mod. Phys. E} \textbf{\bibinfo{volume}{22}},
  \bibinfo{pages}{1330016} (\bibinfo{year}{2013}).

\bibitem{suhonen2007:nucleons-nucleus}
\bibinfo{author}{\bibfnamefont{J.}~\bibnamefont{Suhonen}},
  \emph{\bibinfo{title}{From Nucleons to Nucleus}}
  (\bibinfo{publisher}{Springer-Verlag}, \bibinfo{address}{Berlin},
  \bibinfo{year}{2007}).

\bibitem{weniger1985:fourier-plane-wave}
\bibinfo{author}{\bibfnamefont{E.~J.} \bibnamefont{Weniger}},
  \bibinfo{journal}{J. Math. Phys.} \textbf{\bibinfo{volume}{26}},
  \bibinfo{pages}{276} (\bibinfo{year}{1985}).

\bibitem{shull1955-continuum}
\bibinfo{author}{\bibfnamefont{H.}~\bibnamefont{Shull}} \bibnamefont{and}
  \bibinfo{author}{\bibfnamefont{P.-O.} \bibnamefont{L{\"o}wdin}},
  \bibinfo{journal}{J. Chem. Phys.} \textbf{\bibinfo{volume}{23}},
  \bibinfo{pages}{1362} (\bibinfo{year}{1955}).

\bibitem{rotenberg1962:sturmian-scatt}
\bibinfo{author}{\bibfnamefont{M.}~\bibnamefont{Rotenberg}},
  \bibinfo{journal}{Ann. Phys. (N.Y.)} \textbf{\bibinfo{volume}{19}},
  \bibinfo{pages}{262} (\bibinfo{year}{1962}).

\bibitem{rotenberg1970:sturmian-scatt}
\bibinfo{author}{\bibfnamefont{M.}~\bibnamefont{Rotenberg}},
  \bibinfo{journal}{Adv. At. Mol. Phys.} \textbf{\bibinfo{volume}{6}},
  \bibinfo{pages}{233} (\bibinfo{year}{1970}).

\bibitem{jacobs1986:heavy-quark-sturmian}
\bibinfo{author}{\bibfnamefont{S.}~\bibnamefont{Jacobs}},
  \bibinfo{author}{\bibfnamefont{M.~G.} \bibnamefont{Olsson}},
  \bibnamefont{and} \bibinfo{author}{\bibfnamefont{C.}~\bibnamefont{Suchyta},
  \bibfnamefont{III}}, \bibinfo{journal}{Phys. Rev. D}
  \textbf{\bibinfo{volume}{33}}, \bibinfo{pages}{3338} (\bibinfo{year}{1986}).

\bibitem{keister1997:on-basis}
\bibinfo{author}{\bibfnamefont{B.~D.} \bibnamefont{Keister}} \bibnamefont{and}
  \bibinfo{author}{\bibfnamefont{W.~N.} \bibnamefont{Polyzou}},
  \bibinfo{journal}{J. Comput. Phys.} \textbf{\bibinfo{volume}{134}},
  \bibinfo{pages}{231} (\bibinfo{year}{1997}).

\bibitem{pervin2005:diss}
\bibinfo{author}{\bibfnamefont{M.}~\bibnamefont{Pervin}},
  \emph{\bibinfo{title}{Semileptonic decay of heavy baryons in a constituent
  quark model}}, Ph.D. thesis, \bibinfo{school}{Florida State University}
  (\bibinfo{year}{2005}).

\bibitem{smith2012:fishbone-sturmian}
\bibinfo{author}{\bibfnamefont{E.~S.~R.} \bibnamefont{Woodhouse}}
  \bibnamefont{and} \bibinfo{author}{\bibfnamefont{Z.}~\bibnamefont{Papp}},
  \bibinfo{journal}{Phys. Rev. C} \textbf{\bibinfo{volume}{86}},
  \bibinfo{pages}{067001} (\bibinfo{year}{2012}).

\bibitem{caprio2012:csbasis}
\bibinfo{author}{\bibfnamefont{M.~A.} \bibnamefont{Caprio}},
  \bibinfo{author}{\bibfnamefont{P.}~\bibnamefont{Maris}}, \bibnamefont{and}
  \bibinfo{author}{\bibfnamefont{J.~P.} \bibnamefont{Vary}},
  \bibinfo{journal}{Phys. Rev. C} \textbf{\bibinfo{volume}{86}},
  \bibinfo{pages}{034312} (\bibinfo{year}{2012}).

\bibitem{coon2013:ncsm-extrapolation-ntse13-DUMMY}
\bibinfo{author}{\bibfnamefont{S.}~\bibnamefont{Coon}} \bibnamefont{and}
  \bibinfo{author}{\bibfnamefont{M.~K.~G.} \bibnamefont{Kruse}}, in
  \emph{\bibinfo{booktitle}{Proceedings of the International Conference Nuclear
  Theory in the Supercomputing Era 2013}}, edited by
  \bibinfo{editor}{\bibfnamefont{A.~M.} \bibnamefont{Shirokov}}
  \bibnamefont{and} \bibinfo{editor}{\bibfnamefont{A.~I.} \bibnamefont{Mazur}}
  (\bibinfo{publisher}{Pacific National University, Khabarovsk, Russia},
  \bibinfo{year}{2014}), p. \bibinfo{pages}{314}.

\bibitem{furnstahl2014:ir-expansion}
\bibinfo{author}{\bibfnamefont{R.~J.} \bibnamefont{Furnstahl}},
  \bibinfo{author}{\bibfnamefont{S.~N.} \bibnamefont{More}}, \bibnamefont{and}
  \bibinfo{author}{\bibfnamefont{T.}~\bibnamefont{Papenbrock}},
  \bibinfo{journal}{Phys. Rev. C} \textbf{\bibinfo{volume}{89}},
  \bibinfo{pages}{044301} (\bibinfo{year}{2014}).

\bibitem{caprio2013:cshalo-ntse13}
\bibinfo{author}{\bibfnamefont{M.~A.} \bibnamefont{Caprio}},
  \bibinfo{author}{\bibfnamefont{P.}~\bibnamefont{Maris}}, \bibnamefont{and}
  \bibinfo{author}{\bibfnamefont{J.~P.} \bibnamefont{Vary}}, in
  \emph{\bibinfo{booktitle}{Proceedings of the International Conference Nuclear
  Theory in the Supercomputing Era 2013}}, edited by
  \bibinfo{editor}{\bibfnamefont{A.~M.} \bibnamefont{Shirokov}}
  \bibnamefont{and} \bibinfo{editor}{\bibfnamefont{A.~I.} \bibnamefont{Mazur}}
  (\bibinfo{publisher}{Pacific National University, Khabarovsk, Russia},
  \bibinfo{year}{2014}), p. \bibinfo{pages}{325}.

\bibitem{moshinsky1996:oscillator}
\bibinfo{author}{\bibfnamefont{M.}~\bibnamefont{Moshinsky}} \bibnamefont{and}
  \bibinfo{author}{\bibfnamefont{Y.~F.} \bibnamefont{Smirnov}},
  \emph{\bibinfo{title}{The Harmonic Oscillator in Modern Physics}}
  (\bibinfo{publisher}{Harwood Academic Publishers},
  \bibinfo{address}{Amsterdam}, \bibinfo{year}{1996}).

\bibitem{gloeckner1974:spurious-com}
\bibinfo{author}{\bibfnamefont{D.~H.} \bibnamefont{Gloeckner}}
  \bibnamefont{and} \bibinfo{author}{\bibfnamefont{R.~D.}
  \bibnamefont{Lawson}}, \bibinfo{journal}{Phys. Lett. B}
  \textbf{\bibinfo{volume}{53}}, \bibinfo{pages}{313} (\bibinfo{year}{1974}).

\bibitem{hagen2006:gdm-realistic}
\bibinfo{author}{\bibfnamefont{G.}~\bibnamefont{Hagen}},
  \bibinfo{author}{\bibfnamefont{M.}~\bibnamefont{Hjorth-Jensen}},
  \bibnamefont{and} \bibinfo{author}{\bibfnamefont{N.}~\bibnamefont{Michel1}},
  \bibinfo{journal}{Phys. Rev. C} \textbf{\bibinfo{volume}{73}},
  \bibinfo{pages}{064307} (\bibinfo{year}{2006}).

\bibitem{lu2013:laser-neutron-rich}
\bibinfo{author}{\bibfnamefont{Z.-T.} \bibnamefont{Lu}},
  \bibinfo{author}{\bibfnamefont{P.}~\bibnamefont{Mueller}},
  \bibinfo{author}{\bibfnamefont{G.~W.~F.} \bibnamefont{Drake}},
  \bibinfo{author}{\bibfnamefont{W.}~\bibnamefont{N{\"o}rtersh{\"a}user}},
  \bibinfo{author}{\bibfnamefont{S.~C.} \bibnamefont{Pieper}},
  \bibnamefont{and} \bibinfo{author}{\bibfnamefont{Z.-C.} \bibnamefont{Yan}},
  \bibinfo{journal}{Rev. Mod. Phys.} \textbf{\bibinfo{volume}{85}},
  \bibinfo{pages}{1383} (\bibinfo{year}{2013}).

\bibitem{sick2008:escatt-eval}
\bibinfo{author}{\bibfnamefont{I.}~\bibnamefont{Sick}}, \bibinfo{journal}{Phys.
  Rev. C} \textbf{\bibinfo{volume}{77}}, \bibinfo{pages}{041302(R)}
  (\bibinfo{year}{2008}).

\bibitem{wang2004:6he-radius-laser}
\bibinfo{author}{\bibfnamefont{L.-B.} \bibnamefont{Wang}},
  \bibinfo{author}{\bibfnamefont{P.}~\bibnamefont{Mueller}},
  \bibinfo{author}{\bibfnamefont{K.}~\bibnamefont{Bailey}},
  \bibinfo{author}{\bibfnamefont{G.~W.~F.} \bibnamefont{Drake}},
  \bibinfo{author}{\bibfnamefont{J.~P.} \bibnamefont{Greene}},
  \bibinfo{author}{\bibfnamefont{D.}~\bibnamefont{Henderson}},
  \bibinfo{author}{\bibfnamefont{R.~J.} \bibnamefont{Holt}},
  \bibinfo{author}{\bibfnamefont{R.~V.~F.} \bibnamefont{Janssens}},
  \bibinfo{author}{\bibfnamefont{C.~L.} \bibnamefont{Jiang}},
  \bibinfo{author}{\bibfnamefont{Z.-T.} \bibnamefont{Lu}},
  \bibinfo{author}{\bibfnamefont{T.~P.} \bibnamefont{O'Connor}},
  \bibinfo{author}{\bibfnamefont{R.~C.} \bibnamefont{Pardo}},
  \bibinfo{author}{\bibfnamefont{K.~E.} \bibnamefont{Rehm}},
  \bibinfo{author}{\bibfnamefont{J.~P.} \bibnamefont{Schiffer}},
  \bibnamefont{and} \bibinfo{author}{\bibfnamefont{X.~D.} \bibnamefont{Tang}},
  \bibinfo{journal}{Phys. Rev. Lett.} \textbf{\bibinfo{volume}{93}},
  \bibinfo{pages}{142501} (\bibinfo{year}{2004}).

\bibitem{mueller2007:8he-radius-laser}
\bibinfo{author}{\bibfnamefont{P.}~\bibnamefont{Mueller}},
  \bibinfo{author}{\bibfnamefont{I.~A.} \bibnamefont{Sulai}},
  \bibinfo{author}{\bibfnamefont{A.~C.~C.} \bibnamefont{Villari}},
  \bibinfo{author}{\bibfnamefont{J.~A.} \bibnamefont{Alc\'antara-N\'u\~nez}},
  \bibinfo{author}{\bibfnamefont{R.}~\bibnamefont{Alves-Cond\'e}},
  \bibinfo{author}{\bibfnamefont{K.}~\bibnamefont{Bailey}},
  \bibinfo{author}{\bibfnamefont{G.~W.~F.} \bibnamefont{Drake}},
  \bibinfo{author}{\bibfnamefont{M.}~\bibnamefont{Dubois}},
  \bibinfo{author}{\bibfnamefont{C.}~\bibnamefont{El\'eon}},
  \bibinfo{author}{\bibfnamefont{G.}~\bibnamefont{Gaubert}},
  \bibinfo{author}{\bibfnamefont{R.~J.} \bibnamefont{Holt}},
  \bibinfo{author}{\bibfnamefont{R.~V.~F.} \bibnamefont{Janssens}},
  \bibinfo{author}{\bibfnamefont{N.}~\bibnamefont{Lecesne}},
  \bibinfo{author}{\bibfnamefont{Z.-T.} \bibnamefont{Lu}},
  \bibinfo{author}{\bibfnamefont{T.~P.} \bibnamefont{O'Connor}},
  \bibinfo{author}{\bibfnamefont{M.-G.} \bibnamefont{Saint-Laurent}},
  \bibinfo{author}{\bibfnamefont{J.-C.} \bibnamefont{Thomas}},
  \bibnamefont{and} \bibinfo{author}{\bibfnamefont{L.-B.} \bibnamefont{Wang}},
  \bibinfo{journal}{Phys. Rev. Lett.} \textbf{\bibinfo{volume}{99}},
  \bibinfo{pages}{252501} (\bibinfo{year}{2007}).

\bibitem{brodeur2013:6he-8he-mass}
\bibinfo{author}{\bibfnamefont{M.}~\bibnamefont{Brodeur}},
  \bibinfo{author}{\bibfnamefont{T.}~\bibnamefont{Brunner}},
  \bibinfo{author}{\bibfnamefont{C.}~\bibnamefont{Champagne}},
  \bibinfo{author}{\bibfnamefont{S.}~\bibnamefont{Ettenauer}},
  \bibinfo{author}{\bibfnamefont{M.~J.} \bibnamefont{Smith}},
  \bibinfo{author}{\bibfnamefont{A.}~\bibnamefont{Lapierre}},
  \bibinfo{author}{\bibfnamefont{R.}~\bibnamefont{Ringle}},
  \bibinfo{author}{\bibfnamefont{V.~L.} \bibnamefont{Ryjkov}},
  \bibinfo{author}{\bibfnamefont{S.}~\bibnamefont{Bacca}},
  \bibinfo{author}{\bibfnamefont{P.}~\bibnamefont{Delheij}},
  \bibinfo{author}{\bibfnamefont{G.~W.~F.} \bibnamefont{Drake}},
  \bibinfo{author}{\bibfnamefont{D.}~\bibnamefont{Lunney}},
  \bibinfo{author}{\bibfnamefont{A.}~\bibnamefont{Schwenk}}, \bibnamefont{and}
  \bibinfo{author}{\bibfnamefont{J.}~\bibnamefont{Dilling}},
  \bibinfo{journal}{Phys. Rev. Lett.} \textbf{\bibinfo{volume}{108}},
  \bibinfo{pages}{052504} (\bibinfo{year}{2013}).

\bibitem{friar1997:charge-radius-correction}
\bibinfo{author}{\bibfnamefont{J.~L.} \bibnamefont{Friar}},
  \bibinfo{author}{\bibfnamefont{J.}~\bibnamefont{Martorell}},
  \bibnamefont{and} \bibinfo{author}{\bibfnamefont{D.~W.~L.}
  \bibnamefont{Sprung}}, \bibinfo{journal}{Phys. Rev. A}
  \textbf{\bibinfo{volume}{56}}, \bibinfo{pages}{4579} (\bibinfo{year}{1997}).

\bibitem{tanihata1985:radii-he}
\bibinfo{author}{\bibfnamefont{I.}~\bibnamefont{Tanihata}},
  \bibinfo{author}{\bibfnamefont{H.}~\bibnamefont{Hamagaki}},
  \bibinfo{author}{\bibfnamefont{O.}~\bibnamefont{Hashimoto}},
  \bibinfo{author}{\bibfnamefont{S.}~\bibnamefont{Nagamiya}},
  \bibinfo{author}{\bibfnamefont{Y.}~\bibnamefont{Shida}},
  \bibinfo{author}{\bibfnamefont{N.}~\bibnamefont{Yoshikawa}},
  \bibinfo{author}{\bibfnamefont{O.}~\bibnamefont{Yamakawa}},
  \bibinfo{author}{\bibfnamefont{K.}~\bibnamefont{Sugimoto}},
  \bibinfo{author}{\bibfnamefont{T.}~\bibnamefont{Kobayashi}},
  \bibinfo{author}{\bibfnamefont{D.~E.} \bibnamefont{Greiner}},
  \bibinfo{author}{\bibfnamefont{N.}~\bibnamefont{Takahashi}},
  \bibnamefont{and} \bibinfo{author}{\bibfnamefont{Y.}~\bibnamefont{Nojiri}},
  \bibinfo{journal}{Phys. Lett. B} \textbf{\bibinfo{volume}{160}},
  \bibinfo{pages}{380} (\bibinfo{year}{1985}).

\bibitem{alkhazov2002:elastic-halo-radii}
\bibinfo{author}{\bibfnamefont{G.}~\bibnamefont{Alkhazov}},
  \bibinfo{author}{\bibfnamefont{A.}~\bibnamefont{Dobrovolsky}},
  \bibinfo{author}{\bibfnamefont{P.}~\bibnamefont{Egelhof}},
  \bibinfo{author}{\bibfnamefont{H.}~\bibnamefont{Geissel}},
  \bibinfo{author}{\bibfnamefont{H.}~\bibnamefont{Irnich}},
  \bibinfo{author}{\bibfnamefont{A.}~\bibnamefont{Khanzadeev}},
  \bibinfo{author}{\bibfnamefont{G.}~\bibnamefont{Korolev}},
  \bibinfo{author}{\bibfnamefont{A.}~\bibnamefont{Lobodenko}},
  \bibinfo{author}{\bibfnamefont{G.}~\bibnamefont{Münzenberg}},
  \bibinfo{author}{\bibfnamefont{M.}~\bibnamefont{Mutterer}},
  \bibinfo{author}{\bibfnamefont{S.}~\bibnamefont{Neumaier}},
  \bibinfo{author}{\bibfnamefont{W.}~\bibnamefont{Schwab}},
  \bibinfo{author}{\bibfnamefont{D.}~\bibnamefont{Seliverstov}},
  \bibinfo{author}{\bibfnamefont{T.}~\bibnamefont{Suzuki}}, \bibnamefont{and}
  \bibinfo{author}{\bibfnamefont{A.}~\bibnamefont{Vorobyov}},
  \bibinfo{journal}{Nucl. Phys. A} \textbf{\bibinfo{volume}{712}},
  \bibinfo{pages}{269} (\bibinfo{year}{2002}).

\bibitem{tanihata1988:radii-be-b-halo}
\bibinfo{author}{\bibfnamefont{I.}~\bibnamefont{Tanihata}},
  \bibinfo{author}{\bibfnamefont{T.}~\bibnamefont{Kobayashi}},
  \bibinfo{author}{\bibfnamefont{O.}~\bibnamefont{Yamakawa}},
  \bibinfo{author}{\bibfnamefont{S.}~\bibnamefont{Shimoura}},
  \bibinfo{author}{\bibfnamefont{K.}~\bibnamefont{Ekuni}},
  \bibinfo{author}{\bibfnamefont{K.}~\bibnamefont{Sugimoto}},
  \bibinfo{author}{\bibfnamefont{N.}~\bibnamefont{Takahashi}},
  \bibinfo{author}{\bibfnamefont{T.}~\bibnamefont{Shimoda}}, \bibnamefont{and}
  \bibinfo{author}{\bibfnamefont{H.}~\bibnamefont{Sato}},
  \bibinfo{journal}{Phys. Lett. B} \textbf{\bibinfo{volume}{206}},
  \bibinfo{pages}{592} (\bibinfo{year}{1988}).

\bibitem{tanihata1992:neutron-skins}
\bibinfo{author}{\bibfnamefont{I.}~\bibnamefont{Tanihata}},
  \bibinfo{author}{\bibfnamefont{D.}~\bibnamefont{Hirata}},
  \bibinfo{author}{\bibfnamefont{T.}~\bibnamefont{Kobayashi}},
  \bibinfo{author}{\bibfnamefont{S.}~\bibnamefont{Shirnoura}},
  \bibinfo{author}{\bibfnamefont{K.}~\bibnamefont{Sugimoto}}, \bibnamefont{and}
  \bibinfo{author}{\bibfnamefont{H.}~\bibnamefont{Toki}},
  \bibinfo{journal}{Phys. Lett. B} \textbf{\bibinfo{volume}{289}},
  \bibinfo{pages}{261} (\bibinfo{year}{1992}).

\bibitem{alkhalili2003:inelastic-halo-radii}
\bibinfo{author}{\bibfnamefont{J.~S.} \bibnamefont{Al-Khalili}},
  \bibinfo{author}{\bibfnamefont{J.~A.} \bibnamefont{Tostevin}},
  \bibnamefont{and} \bibinfo{author}{\bibfnamefont{I.~J.}
  \bibnamefont{Thompson}}, \bibinfo{journal}{Phys. Rev. C}
  \textbf{\bibinfo{volume}{54}}, \bibinfo{pages}{1843} (\bibinfo{year}{1996}).

\bibitem{shirokov2007:nn-jisp16}
\bibinfo{author}{\bibfnamefont{A.~M.} \bibnamefont{Shirokov}},
  \bibinfo{author}{\bibfnamefont{J.~P.} \bibnamefont{Vary}},
  \bibinfo{author}{\bibfnamefont{A.~I.} \bibnamefont{Mazur}}, \bibnamefont{and}
  \bibinfo{author}{\bibfnamefont{T.~A.} \bibnamefont{Weber}},
  \bibinfo{journal}{Phys. Lett. B} \textbf{\bibinfo{volume}{644}},
  \bibinfo{pages}{33} (\bibinfo{year}{2007}).

\bibitem{sternberg2008:ncsm-mfdn-sc08}
\bibinfo{author}{\bibfnamefont{P.}~\bibnamefont{Sternberg}},
  \bibinfo{author}{\bibfnamefont{E.~G.} \bibnamefont{Ng}},
  \bibinfo{author}{\bibfnamefont{C.}~\bibnamefont{Yang}},
  \bibinfo{author}{\bibfnamefont{P.}~\bibnamefont{Maris}},
  \bibinfo{author}{\bibfnamefont{J.~P.} \bibnamefont{Vary}},
  \bibinfo{author}{\bibfnamefont{M.}~\bibnamefont{Sosonkina}},
  \bibnamefont{and} \bibinfo{author}{\bibfnamefont{H.~V.} \bibnamefont{Le}}, in
  \emph{\bibinfo{booktitle}{SC '08: Proceedings of the 2008 ACM/IEEE Conference
  on Supercomputing}} (\bibinfo{publisher}{IEEE Press},
  \bibinfo{address}{Piscataway, NJ}, \bibinfo{year}{2008}),
  \bibinfo{note}{{A}rticle No. 15}.

\bibitem{maris2010:ncsm-mfdn-iccs10}
\bibinfo{author}{\bibfnamefont{P.}~\bibnamefont{Maris}},
  \bibinfo{author}{\bibfnamefont{M.}~\bibnamefont{Sosonkina}},
  \bibinfo{author}{\bibfnamefont{J.~P.} \bibnamefont{Vary}},
  \bibinfo{author}{\bibfnamefont{E.}~\bibnamefont{Ng}}, \bibnamefont{and}
  \bibinfo{author}{\bibfnamefont{C.}~\bibnamefont{Yang}},
  \bibinfo{journal}{Procedia Comput. Sci.} \textbf{\bibinfo{volume}{1}},
  \bibinfo{pages}{97} (\bibinfo{year}{2010}).

\bibitem{aktulga2013:mfdn-ONLINE}
\bibinfo{author}{\bibfnamefont{H.~M.} \bibnamefont{Aktulga}},
  \bibinfo{author}{\bibfnamefont{C.}~\bibnamefont{Yang}},
  \bibinfo{author}{\bibfnamefont{E.~G.} \bibnamefont{Ng}},
  \bibinfo{author}{\bibfnamefont{P.}~\bibnamefont{Maris}}, \bibnamefont{and}
  \bibinfo{author}{\bibfnamefont{J.~P.} \bibnamefont{Vary}},
  \bibinfo{journal}{Concurrency Computat.: Pract. Exper.}
  (\bibinfo{year}{2013}), \bibinfo{note}{{DOI}: 10.1002/cpe.3129}.

\bibitem{shirokov2014:jisp16-binding}
\bibinfo{author}{\bibfnamefont{A.~M.} \bibnamefont{Shirokov}},
  \bibinfo{author}{\bibfnamefont{V.~A.} \bibnamefont{Kulikov}},
  \bibinfo{author}{\bibfnamefont{P.}~\bibnamefont{Maris}}, \bibnamefont{and}
  \bibinfo{author}{\bibfnamefont{J.~P.} \bibnamefont{Vary}}, in
  \emph{\bibinfo{booktitle}{$NN$ and $3N$ Interactions}}, edited by
  \bibinfo{editor}{\bibfnamefont{L.~D.} \bibnamefont{Blokhintsev}}
  \bibnamefont{and} \bibinfo{editor}{\bibfnamefont{I.~I.}
  \bibnamefont{Strakovsky}} (\bibinfo{publisher}{Nova Science},
  \bibinfo{address}{Hauppauge, N.Y.}, \bibinfo{year}{2014}),
  Chap.~\bibinfo{chapter}{8}.

\bibitem{constantinou-IP}
\bibinfo{author}{\bibnamefont{\mbox{Ch}. Constantinou}}
  \bibnamefont{\emph{et~al.}} (\bibinfo{year}{in preparation}).

\bibitem{furnstahl-PREPRINT:ir-extrapolation-cc-16o}
\bibinfo{author}{\bibfnamefont{R.~J.} \bibnamefont{Furnstahl}},
  \bibinfo{author}{\bibfnamefont{G.}~\bibnamefont{Hagen}},
  \bibinfo{author}{\bibfnamefont{T.}~\bibnamefont{Papenbrock}},
  \bibnamefont{and} \bibinfo{author}{\bibfnamefont{K.~A.} \bibnamefont{Wendt}},
  \bibinfo{journal}{J. Phys. G}  (\bibinfo{year}{submitted}),
  \eprint{1203.2515}.

\bibitem{negoita2010:diss}
\bibinfo{author}{\bibfnamefont{G.~A.} \bibnamefont{Negoita}},
  \emph{\bibinfo{title}{\textit{Ab initio} nuclear structure theory}}, Ph.D.
  thesis, \bibinfo{school}{Iowa State University} (\bibinfo{year}{2010}).

\bibitem{negoita-IP}
\bibinfo{author}{\bibfnamefont{G.~A.} \bibnamefont{Negoita}}
  \bibnamefont{\emph{et~al.}} (\bibinfo{year}{in preparation}).

\bibitem{bacca2012:6he-hyperspherical}
\bibinfo{author}{\bibfnamefont{S.}~\bibnamefont{Bacca}},
  \bibinfo{author}{\bibfnamefont{N.}~\bibnamefont{Barnea}}, \bibnamefont{and}
  \bibinfo{author}{\bibfnamefont{A.}~\bibnamefont{Schwenk}},
  \bibinfo{journal}{Phys. Rev. C} \textbf{\bibinfo{volume}{86}},
  \bibinfo{pages}{034321} (\bibinfo{year}{2012}).

\bibitem{quaglioni2013:ncsm-rgm-cluster-6he}
\bibinfo{author}{\bibfnamefont{S.}~\bibnamefont{Quaglioni}},
  \bibinfo{author}{\bibfnamefont{C.}~\bibnamefont{Romero-Redondo}},
  \bibnamefont{and}
  \bibinfo{author}{\bibfnamefont{P.}~\bibnamefont{Navr{\'a}til}},
  \bibinfo{journal}{Phys. Rev. C} \textbf{\bibinfo{volume}{88}},
  \bibinfo{pages}{034320} (\bibinfo{year}{2013}).

\bibitem{helgaker2000:electron-structure}
\bibinfo{author}{\bibfnamefont{T.}~\bibnamefont{Helgaker}},
  \bibinfo{author}{\bibfnamefont{P.}~\bibnamefont{J{\o}rgensen}},
  \bibnamefont{and} \bibinfo{author}{\bibfnamefont{J.}~\bibnamefont{Olsen}},
  \emph{\bibinfo{title}{Molecular Electron-Structure Theory}}
  (\bibinfo{publisher}{Wiley}, \bibinfo{address}{Chichester},
  \bibinfo{year}{2000}).

\end{thebibliography}
\end{document}